\documentclass[debug]{rmaa}

\usepackage{paralist}

\usepackage{psfrag,color}
\usepackage{diagbox}
\usepackage{adjustbox}
\usepackage{natbib}
\usepackage{pdflscape}
\usepackage[latin1]{inputenc}

\title{X-ray analysis of Seyfert 1 galaxies with optical polarization: a test for unification models\footnote{Based on observations obtained with XMM-Newton, an ESA science mission with instruments and contributions directly funded by ESA Member States and NASA}} 

\author{
  Gudi\~no, M.,\altaffilmark{1} 
  Jim\'enez-Bail\'on E.,\altaffilmark{1,2}
  Longinotti, A.L.,\altaffilmark{1}
  Guainazzi, M., \altaffilmark{3}
  Cervi\~no, M., \altaffilmark{4}
  A.C. Robleto-Or\'us.\altaffilmark{1}}
       
\altaffiltext{1}{Instituto de Astronom\'ia, Universidad Nacional Aut\'onoma de M\'exico.}
\altaffiltext{2}{Quasar Science Resource S.L. for the European Space Agency (ESA), European Space Astronomy Centre (ESAC).}
\altaffiltext{3}{ESA, European Space Research and Technology Centre (ESTEC).}
\altaffiltext{4}{Centro de Astrobiolog\'ia (CAB), CSIC-INTA.}

\shortauthor{Gudi\~no, M. et al.}
\shorttitle{X-ray view of polarized Seyfert galaxies.}

\fulladdresses{
\item Instituto de Astronom\'ia, Universidad Nacional Aut\'onoma de M\'exico, AP 70-264, Ciudad de M\'exico, CDMX 04510, Mexico. (megudino@astro.unam.mx).
\item Quasar Science Resource S.L. for the European Space Agency (ESA), European Space Astronomy Centre (ESAC), Camino Bajo del Castillo s/n, 28692 Villanueva de la Ca\~nada, Madrid, Spain.
\item European Space Agency, European Space Research and Technology Centre (ESTEC), Keplerlaan 1, 2201 AZ Noordwijk, The Netherlands.
\item Centro de Astrobiolog\'ia (CAB), CSIC-INTA, Campus ESAC, camino bajo del Castillo s/n, Villafranca del Castillo, Villanueva de la Ca\~nada, 28692  Madrid, Spain.}

\listofauthors{Gudi\~no, M., Longinotti, A.L., Jim\'enez-Bail\'on, E., Guainazzi, M., Cervi\~no, M., \& A.C. Robleto-Or\'us.}
\indexauthor{Gudi\~no, M.}
\indexauthor{Longinotti, A.L.}
\indexauthor{Jim\'enez-Bail\'on, E.}
\indexauthor{Guainazzi, M.}
\indexauthor{Cervi\~no, M.}
\indexauthor{A.C. Robleto-Or\'us.}

\abstract{In accordance with the AGN Unified Model, observed polarization can be related to the orientation of the line of sight with respect to the torus. AGN X-ray emission arises from the central region and carries the imprints of the obscuring material. We aim to test a unified scheme based on optical polarization using X-ray absorption. Using the XMM-Newton data of 19, optically polarized Seyfert 1 sources, we developed a systematic analysis by fitting a baseline model to test the presence of X-ray neutral or ionized (warm) absorption. We find that 100\% of the polar-polarized sources show the presence of absorption, with 70\% favoring the presence of a warm absorber. In contrast, the equatorial-polarized sources show a fraction of absorbed spectra of 75\%, with only 50\% consistent with the presence of a warm absorber.}

\resumen{De acuerdo con el Modelo Unificado AGN, la polarizaci\'on observada se puede relacionar con la orientaci\'on de la l\'inea de visi\'on con respecto al toro. La emisi\'on de rayos X de un AGN surge de la regi\'on central y lleva las huellas del material oscurecedor. Nuestro objetivo es probar un esquema unificado basado en polarizaci\'on \'optica mediante absorci\'on de rayos X. Utilizando los datos XMM-Newton de 19 fuentes Seyfert 1 \'opticamente polarizadas, desarrollamos un an\'alisis sistem\'atico para ajustar un modelo de referencia y probar la presencia de absorci\'on neutra o ionizada. Encontramos que el 100\% de las fuentes con polarizaci\'on polar muestran la presencia de absorci\'on, con el 70\% favoreciendo la presencia de un absorbedor tibio. Por el contrario, las fuentes con polarizaci\'on ecuatorial muestran una fracci\'on de espectros absorbidos del 75\%, y s\'olo el 50\% es consistente con la presencia de un absorbedor tibio.}

\addkeyword{galaxies: active}
\addkeyword{galaxies: nuclei}
\addkeyword{galaxies: Seyfert}
\addkeyword{X-rays: galaxies}

\begin{document}
\maketitle
\section{Introduction}\label{sec:intro}

A galaxy is said to host an active nucleus (AGN) when it exhibits a highly luminous central region, comparable to or even brighter than the integrated light of the stars in the galaxy, with a luminosity extending throughout the entire electromagnetic spectrum. The AGN emission arises from a central compact region that consists of a supermassive black hole (SMBH) surrounded by an accretion disk. A characteristic feature present in some AGN spectra are optical broad emission lines, \citep[full width half maximum (FWHM) $\sim 1000-20000\ \mathrm{km\ s^{-1}}$,][]{N2015}, from a high-density gas region that is excited and ionized by the central engine, located at $0.1-1\ \mathrm{pc}$ from it, \citep{K2005}, the Broad Line Region (BLR). The observation of optical narrow lines (FWHM $\sim 300-1000\ \mathrm{km\ s^{-1}}$) indicates the presence of another gas region of lower density and ionization known as the Narrow Line Region (NLR), this region extends to a scale of $\sim 100\ \mathrm{pc}$ up to $1000\ \mathrm{pc}$, \citep{N2015, Volk2012}.

The absence of broad emission lines in some AGN spectra is explained by postulating that an optically thick toroidal structure made of neutral gas and dust extending up to $\sim 10\ \mathrm{pc}$, surrounds the BLR and central engine, with the ultimate effect of obscuring the direct AGN emission from the observer view. The Unified Model, \citep{AM1985,RA1993}, states that different features observed in AGN spectra depend on the orientation of this obscuring torus relative to our line-of-sight. From this orientation effect, we distinguish two main AGN types. Type 1 are looked directly into the central region, showing both narrow and broad optical emission lines. Type 2 are looked through the torus, so the central region and BLR get obscured and only narrow emission lines are detected. A more detailed classification is provided by looking to specific spectral lines, e.g. a Narrow Line Seyfert 1 (NLSy1) source show a much narrower H$\beta$ line compared to classical type 1 AGN (FWHM $< 2000$ kms$^{-1}$) and an unusually strong $\mathrm{Fe} \sc{II}$ line \citep{O1985, Ko2008}, thus providing a wide range of AGN types. In the case of this work, we will adopt the simplest classification with the aim to relate it to optical polarization.

We focus our work on optically polarized type 1 Seyfert galaxies (Sy). The findings of polarized broad emission lines in type 2 Sy, e.g. NGC 1068 \citep{MGM1991}, suggested that the BLR is present but obscured from the line-of sight, lending support to the Unification Model. In general, the polarization position angle (PA) of type 2 Sy is found to be perpendicular to the main axis of the system, i.e. the rotation axis of the accretion disk. In this case, the polar-polarization region, {\it PL-pol}, corresponds to the well-established AGN ionization cones that trace the kpc-scale NLR, \citet[][hereafter S02, S04]{Smith2002, Smith2004}. In the framework envisaged by Smith, a second scattering region is postulated to account for the observations on polarized type 1 Sy, \citet[][hereafter S05]{Smith20042}, in this case the  polarization position angle is parallel to the main axis of the system. This so-called  equatorial-polarization region, {\it EQ-pol}, is co-planar to the accretion disk. However, we note that additional contributions from the nuclear components may determine the type of observed polarization, e.g. the accretion disk, as proposed by \citep{Pio2023}, bringing more complexity to the original classification proposed by Smith. As an example, studies report that the optical lines from the {\it EQ-pol} region show the effects of the rotating motion of the scattering region emitting the gas, which are visible in a characteristically S-shaped profile detected in the polarization position angle, \citep{Af2015, Af2019}.

Interestingly, Smith reports a list of type 1 sources with polarization properties similar to those of type 2, i.e. {\it PL-pol}, (S02,S05). Smith proposed a unified scheme addressed to relate these observations with the orientation of the torus: a Sy 1 with {\it PL-pol} represents and intermediate type between a Sy 2 with {\it PL-pol} and a Sy 1 with {\it EQ-pol}. This, in turn, would imprint different observing properties due to the absence ({\it EQ-pol} Sy 1) or presence ({\it PL-pol} Sy 1) of absorbing material co-spatial with the outer layers of the torus on the line of sight. In this framework, the X-ray regime is especially appropriate for assessing the role of the torus absorption effect.

X-ray emission in AGN arises from the up-scattering of UV photons, emitted by the accretion disk, in a surrounding corona of hot electrons, \citep{H1991}. The primary feature of a Seyfert X-ray spectrum is a {\it power-law continuum}, extending up to $100-200\ keV$ and with photon indices varying between $1.4$ and $2.3$, \citep{Pi2005, Rui2017}. At $6.4\ \mathrm{keV}$, we can observe the prominent {\it $K \alpha$ Fe emission line}, \citep{Ri2014}, typically present in all Seyfert galaxies, produced via fluorescence mechanisms by  reprocessing of the primary emission  in circumnuclear neutral material. At energies $<2\ \mathrm{keV}$, the spectrum may exhibit emission that exceeds the flux of the main continuum, called {\it soft-excess}, and sometimes interpreted as thermal radiation from the accretion disk, with black body temperatures in the range of $0.1-1.0\ \mathrm{keV}$, \citep{P2018}. Analogously to the optical, in the X-rays we can distinguish between type 1 and 2 through absorption. This distinction is mainly based on the detection of  Hydrogen column densities that can be as low as $N_H \sim 10^{19}\ \mathrm{cm^{-2}}$, for a type 1 and $N_H \sim 10^{23}\ \mathrm{cm^{-2}}$ or higher in the case of type 2. The main dividing value between of $N_H=10^{22}\ \mathrm{cm^{-2}}$, \citep{Volk2012}. This refers only to absorption by  neutral material, the so called {\it{cold absorber}}. Absorption by ionized material, the {\it{warm absorber}}, can also be present in the line-of-sight, and is generally observed as an outflow (100-2000 $\mathrm{km\ s^{-1}}$) of gas with column densities of $N_H\ \sim10^{21}$ to $10^{22.5}\ \mathrm{cm^{-2}}$ and ionization parameter of $\log\xi=-1$ to $3\ \mathrm{erg\ cm\ s^{-1}}$, \citep{L2021}. 

The study of X-ray absorption can provide an independent test to the polarization unification scheme described above by comparing the X-ray absorption properties, if present, to the known optical polarization characteristics. We remark that although our analysis is specifically addressed to test the presence and nature of X-ray absorption, for the time being, it provides a first approximation point of comparison with the polarization characteristics reported by S02, S04, S05. Further ideas on how to expand it are presented in the Discussion.


We carried out a systematic analysis of the X-ray spectra of 19 Sy 1 galaxies, classified in two sub-samples according to their optical polarization, as detailed in Section 2. This paper is structured as follows: the sample details are described in Section \ref{sect:sample}, followed by a description of our analysis in Section \ref{sect:analysis}. In Sections \ref{sect:results} and \ref{sect:Disc} we present our results and discussion, respectively, finalizing with conclusions in \ref{sect:Conc}. For all our analysis we adopted the standard $\Lambda\ CDM$ cosmology, with the parameters: $H_0=70\ \mathrm{km\ s^{-1}\ Mpc^{-1}}$, $\Omega\Lambda = 0.73$ and $\Omega M = 0.27$. 

\section{XMM-Newton sample and data reduction}\label{sect:sample}

Our sample consists of 19 sources, listed in Table \ref{tab:sample}, selected from the sample of 46 galaxies studied and classified by S02, S04, S05: polar polarization was reported for 11 sources ({\it PL-pol}) whereas 8 sources have equatorial polarization ({\it EQ-pol}). We obtained the {\it EPIC-pn} data at CCD resolution from the {\it XMM-Newton} archive\footnote{http://nxsa.esac.esa.int/nxsa-web/search}. For sources with more than one observation we selected the one with the longest exposure time regardless of the X-ray spectral state for the few sources in our sample with known prominent spectral changes (e.g. NGC 3227 and ESO 323-G077).

To process the data and extract the spectra, we utilized the {\it SAS v19.1}\footnote{User's Guide to the XMM-Newton Science Analysis System, Issue 16.0, 2021 (ESA: XMM-Newton SOC)} software. The {\it EPIC-pn} data was reprocessed with the {\it epproc} task and was filtered for high background events using the standard procedure developed by the {\it XMM-Newton} Science Operations Center (SOC). Utilizing the SAO Image DS9 display \citep{DS9}, we select a circular region that encloses the source, centered at the peak of X-ray emission, with radius ranging from 30 to 68 arcsec, depending on the target. With the {\it evselect} task, we selected the source region and extracted the corresponding spectrum. We used the same task to extract the background spectrum, selected from with a circular region with no contribution from other sources in the CCD. The radius of the background regions vary from 50 to 96 arcsec, depending on the target. Response matrices were generated with the task {\it rmfgen} and {\it arfgen}. The resulting spectra were binned with the task {\it specgroup} in order to obtain 25 counts per energy bin, allowing us to use the $\chi^2$ statistics. We corrected for out-of-time events that may occur due to the readout of the CCD as well as checked for possible pile-up. Moderate percentage of pile-up were found in the following sources: Fairall 51, Mrk 704, Mrk 766, NGC 3227, NGC 4593, that were corrected following the standard procedure suggested by the {\it XMM-Newton} Science Operation Centre (SOC)\footnote{https://www.cosmos.esa.int/web/xmm-newton/sas-thread-epatplot}.

\begin{figure}[!t]\centering
    \includegraphics[width=0.8\columnwidth]{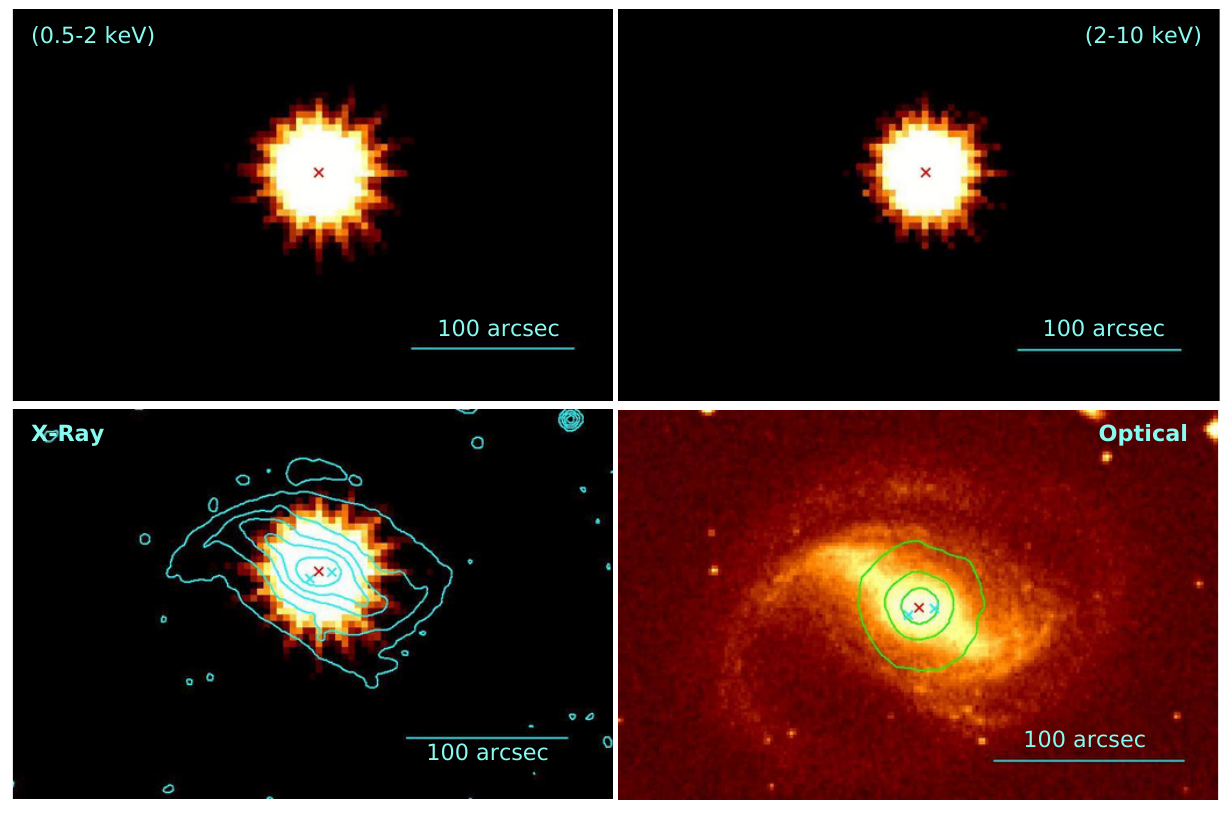}
    \caption{Spatial analysis for NGC 4593. Top panel shows the X-ray images in two different energy ranges: soft X-ray from $0.5-2\ \mathrm{keV}$ and hard X-ray from $2-10\ \mathrm{keV}$. Bottom panel shows a comparison between the X-ray emission ($0.5-10\ \mathrm{keV}$) with optical contours (in green) and an optical image from DSS with X-ray contours (in red). A red cross marks the maximum of X-ray emission, and a cyan cross marks the maximum optical emission.}
    \label{fig:n45spa}
\end{figure}

\begin{table}
\begin{adjustbox}{width=\textwidth}
\begin{tabular}{llrccccr}
\hline
$^{(1)}$ & $^{(2)}$ & $^{(3)}$ & $^{(4)}$ & $^{(5)}$ & $^{(6)}$ & $^{(7)}$ & $^{(8)}$ \\
\hline
Object & RA & Dec & \it{z} & $\mathrm{N^{Gal}_{H}}$ & Type & Obs ID & Exp.\\
  & & & &  $\times 10^{20}\ \mathrm{cm^{-2}}$ & & & ks\\
\hline \\
\textbf{Polar Polarization}\\ \\
Mrk 1218 & 129.546 & +24.8953 & 0.02862 & 3.1 & Sy 1.8 & 0302260201 & 13.9 \\
Mrk 704 & 139.608 & +16.3053 & 0.02923 & 2.7 & Sy 1.2 & 0502091601 & 98.2 \\
Mrk 1239 & 148.080 & -1.6121 & 0.01993 & 4.1 & NLSy1 & 0891070101 & 105.0 \\
NGC 3227 & 155.877 & +19.8651 & 0.00386 & 1.9 & Sy 1.5 & 0782520601 & 107.9 \\
WAS 45 & 181.181 & +31.1772 & 0.02500 & 1.4 & Sy 1.9 & 0601780601 & 39.5 \\
Mrk 766 & 184.611 & +29.8129 & 0.01293 & 1.9 & NLSy1/Sy 1.5 & 0109141301 & 129.9 \\
Mrk 231 & 194.059 & +56.8737 & 0.04217 & 0.9 & Sy 1 & 0770580501 & 26.5 \\
NGC 4593 & 189.914 & -5.3443 & 0.00831 & 1.7 & Sy 1 & 0784740101 & 142.1 \\
ESO 323-G077 & 196.609 & -40.4147 & 0.01501 & 7.7 & Sy 1.2 & 0694170101 & 132.6 \\
IRAS 15091-2107 & 227.999 & -21.3171 & 0.04461 & 7.9 & NLSy1 & 0300240201 & 23.0 \\
Fairall 51 & 281.225 & -62.3648 & 0.01418 & 6.3 & Sy 1 & 0300240401 & 26.9 \\ \\
\textbf{Equatorial Polarization} \\ \\
I Zw1 & 13.396 & +12.6934 & 0.06117 & 4.6 & NLSy1 & 0743050301 & 141.2 \\
Akn 120	& 79.048 & -0.1498 & 0.03271 & 9.9 & Sy 1 & 0721600401 & 133.3 \\
NGC 3783 & 174.757 & -37.7387 & 0.00973 & 0.1 & Sy 1.5 & 0112210201 & 137.8 \\
Mrk 841	& 226.005 & +10.4378 & 0.03642 & 2.0 & Sy 1.5 & 0882130401 & 132.0 \\
Mrk 876	& 243.488 & +65.7193 & 0.12109 & 2.4 & Sy 1 & 0102040601 & 12.8 \\ 
KUV 18217+6419 & 275.489 & +64.3434 & 0.29705 & 3.5 & Sy 1.2 & 0506210101 & 14.3 \\
Mrk 509	& 311.041 & -10.7235 & 0.03440 & 3.9 & Sy 1.5 & 0306090201 & 85.9 \\
Mrk 304 & 334.301 & +14.2391 & 0.06576 & 4.9 & Sy 1 & 0103660301 & 47.3 \\ \\
\hline
\end{tabular}
\end{adjustbox}
\caption{Sample of polarized Seyfert 1 galaxies organized according to their polarization characteristics: polar and equatorial. Columns 2 and 3 report coordinates (RA, Dec) in J2000, column 4 indicates the redshift of each source, taken from the NASA/IPAC Extragalactic Database(NED). Column 5 corresponds to the column density of the Galaxy in the line-of-sight of the source from the HEASARC server. Column 6 indicates AGN classification, where NLSy1 refers to Narrow Line Seyfert 1, from NED. Columns 7 and 8 are the {\it XMM-Newton} observation ID and the net exposure time.}
\label{tab:sample}
\end{table}

\section{X-ray analysis} \label{sect:analysis}

\subsection{Spatial Analysis}

The spatial analysis aims to give a general overview of each source. We used the filtered event files to generate X-ray images of the soft energy band ($0.5-2.0\ \mathrm{keV}$), and the hard energy band ($2.0-10\ \mathrm{keV}$). We also produced a full energy range image $0.5-10\ \mathrm{keV}$ and combined it with the corresponding optical image, taken from the DSS (Digital Sky Survey). We produced combined images by plotting the optical contours in the X-ray image and vice versa. In Fig.\ref{fig:n45spa}, we show the example of the spatial analysis done for NGC 4593. In these images, we can see the point-like X-ray emission of the AGN and a more extended structure in the optical image. With these analysis, we can verify the general characteristic of point-like sources in the X-ray regime. The results of the spatial analysis are presented in section \ref{sect:Rspatial}.\\

\subsection{Spectral Analysis}

For the fitting process, we used the {\it Xspec Spectral Fitting Package v 12.10.1n} in a {\it Python v 3.7.4.} environment: {\it PyXspec}, \citep{PyX}. We considered the Galactic absorption modelled by the Tuebingen-Boulder ISM absorption model, {\it TBabs}, with updated abundances, \citep{W2000}. We used Galactic Hydrogen column values from``nH Column Density Tool" from the HEASARC server, \citep{HI4PI}, see Table \ref{tab:sample}. For all the models that include the redshift parameter, $z$, the value is fixed at the  source redshift. Errors are quoted at $1\sigma$ confidence level or 90 percent for one free parameter.

In order to characterize the X-ray emission, we performed a systematic analysis of the spectra by adding components --either additive or multiplicative-- to a baseline continuum model. As we build a nested model, we can calculate the F-test for additive components, where an F-test $>95\%$ indicates that the new component significantly improves the fit.  For multiplicative components, such as the absorption models, we used the Akaike Information Criterion \citep[AIC,][]{Ak1974}. We first calculate the AIC value, Eq. \ref{eq:one}, where $k$ corresponds to the number of free parameters in each model. The comparison between both models is given by the factor, $F_{\mathrm{AIC}}$, resulting from Eq. \ref{eq:two}, where $\mathrm{AIC_{NEW}}$ refers to the model with the new component. The inverse, $1/F_{\mathrm{AIC}}$, indicates the improvement of the new and more complex model, e.g. \citep{KY2021}.

\begin{equation} \label{eq:one}
    \mathrm{AIC}=2k+\chi^2
\end{equation}
\begin{equation} \label{eq:two}
    F_{\mathrm{AIC}}=\exp ((\mathrm{AIC_{NEW}}-\mathrm{AIC}_{0})/2)
\end{equation}

\subsubsection{The hard band (2-10 keV) spectral range} \label{sect:hb}

The hard band corresponds to the energy range of $2-10\ \mathrm{keV}$. In this range, we fitted the main continuum component with a power law, using the model {\it zpowerlw}, reporting the resulting photon index, $\Gamma$ and normalization in units of $\mathrm{photons\ keV^{-1}\ cm^{-2}\ s^{-1}}$. Subsequently, we tested for the presence of emission lines using the model {\it zgauss}. We only tested for narrow lines, with width fixed to $\sigma=0.05\ \mathrm{keV}$. The most prominent X-ray emission line in AGN is the $\mathrm{Fe\ K\alpha}$ at $6.4\ \mathrm{keV}$, if present, we reported the value of the normalization in units of $\mathrm{photons\ cm^{-2}\ s^{-1}}$. We additionally tested for the presence of narrow emission lines corresponding to ionized Fe: $\mathrm{Fe} \sc{XXV}$ at $6.697\ \mathrm{keV}$ and/or $\mathrm{Fe} \sc{XXVI}$ at $6.966\ \mathrm{keV}$, \citep{Bi2005}. Lastly, we tested for a possible contribution from a cold absorber, fitted using the model {\it zTBabs}. If significant, we reported the column density in units of $10^{22}\ \mathrm{cm^{-2}}$. The results of the hard band fitting are shown in Table \ref{tab:hardband}.

\subsubsection{The full band (0.5-10 keV) spectral range} \label{sect:fe}

Extending our analysis to the full energy band, ($0.5-10\ \mathrm{keV}$), we first applied our hard band baseline model to the EPIC-pn data. This allowed us to verify the presence of additional spectral features such as soft excess and absorption.

In order to fit the soft excess as thermal emission, we assumed a black body spectrum, modeled by the XSPEC component {\it zbbody}, \citep{Si2011, Sc2011, P2018}. We report the black body temperature in $\mathrm{keV}$, and the normalization in units of $L_{39}/[D_{10}(1+z)]^2$; where $L_{39}$ is the luminosity of the source in units of $10^{39}\ \mathrm{erg\ s^{-1}}$ and $D_{10}$ is the distance of the source in units of $10\ \mathrm{pc}$. The results of the baseline model with the addition of the black body fitted to the full energy band are reported in Table \ref{tab:se}.

We now proceed with the absorption test. Absorption is referred to as cold or warm according to the state of the absorbing material, whether it is neutral or ionized. Our aim is to determine if absorption is in fact present in each spectrum and, if so, which component prevails: a cold absorber or a warm absorber. It is worth noting that ionized absorption typically requires a more complex modeling with several layers of gas in different ionization states and column densities, \citep[e.g.][]{GM2013, S2015, Si2018}. All absorption models are multiplicative, so we calculate the AIC and use it to determine which component provides a better fit to the data.

For a cold absorber, we used the model {\it zTBabs}. The fit yields the amount of the line of sight equivalent Hydrogen column in units of $10^{22}\ \mathrm{cm^{-2}}$. For warm absorption, we selected the model {\it zxipcf }\footnote{https://heasarc.gsfc.nasa.gov/xanadu/xspec/models/zxipcf.html}, based on the XSTAR photoionization absorption models, and consider four parameters: the redshift of the source, the column density in units of $10^{22}\ \mathrm{cm^{-2}}$, the covering fraction, $f$, and the ionization parameter, $\log\xi$, where $\xi=L/nr^2$: $\mathrm{L}$ is the X-ray luminosity of the source, $\mathrm{n}$ is the electronic density of the ionized gas and $\mathrm{r^2}$ is the squared distance of the ionized cloud to the source of ionizing radiation. The covering factor, $\mathrm{f}$, is fixed to 1.0, so $1-\mathrm{f}$ represents the portion of the source that is seen directly. As a final test, we explored the possibility that the fits for the sources with warm absorption could be further improved by adding a second warm absorber component, also with $f$ fixed to 1.0. The absorption test results are reported in Table \ref{tab:abstest}. All the spectral analysis results are detailed in section \ref{sect:Rspectral}.

\section{Results} \label{sect:results}

In this section, we present the results of the analysis of our sample that consists of 19 sources of which 11 sources with {\it PL-pol} and 8 with {\it EQ-pol}. We begin with the spatial analysis, that provides a general view of each source. We then detailed the results from the spectral analysis, organized according to the known polarization characteristics. In Section \ref{sect:Disc}, we discuss these results.

\subsection{Spatial analysis}\label{sect:Rspatial}

We detailed the spatial analysis performed on the sources in the sample in section \ref{sect:analysis}. The objects in the sample are all type-1 and intermediate type Seyfert galaxies, see Table \ref{tab:sample}. Consistently with their classification, they all appear as point-like sources in the X-ray energy range images. By tracing the optical contours onto the X-ray image, we verify that the X-ray emission is concentrated in the nuclear region, while the optical structure is extended, showing features such as the spiral arms with stellar formation activity. We can conclude that the sample displays general properties consistent with type 1 Sy galaxies.

\subsection{Spectral analysis} \label{sect:Rspectral}

In order to draw a first approximation description regarding the X-ray properties of our sample, we tested if the spectra can be satisfactorily fitted with our nested model, avoiding a source-by-source detailed modeling. To achieve a general good description of how these sources respond to a first-order broadband AGN emission model, we use the F-test and the AIC for additive and multiplicative respectively. In particular, our aim is to determine if the fit is improved by adding an absorption component and, if significant, whether this absorption is cold or warm. The results of our analysis are listed in Tables \ref{tab:hardband}, \ref{tab:se}, and \ref{tab:abstest}, corresponding to the hard band fits, the soft excess and the absorption test. In Table \ref{tab:abstest} we also include the source luminosity corresponding to the preferred model. The standard threshold for estimating parameter errors in our PyXspec script, and in Xspec in general, is $\chi^2_{\nu}>2$, therefore, we do not report errors for fits with higher values of the $\chi^2$ statistics.

We have organized the description of the results according to the optical polarization classification of the sources, -- {\it PL-pol} and {\it EQ-pol} --. These results will be discussed in section \ref{sect:Disc}. The fits of the entire sample in the full energy band are presented in Appendix A.

\subsubsection{Polar-polarized sources}

The photon index of the hard band power law continuum (see Table \ref{tab:hardband}) ranges from the flatter 0.6 to the steeper 4.3. In particular, for 8 sources the photon index is in a range of $1.4<\Gamma<2.1$ (mean=1.7). Also in the hard band, we tested the presence and significance of emission lines corresponding to $\mathrm{Fe\ K\alpha}$, $\mathrm{Fe \sc{XXV}}$ and $\mathrm{Fe \sc{XXVI}}$. We found significant emission of $\mathrm{Fe\ K\alpha}$ line in 9 of the sources, with only Mrk 1218 and Mrk 231 not showing any emission line. Additionally, NGC 4593 and ESO 323-G077 show the $\mathrm{Fe} \sc{XXVI}$ line, and Mrk 766 shows both $\mathrm{Fe} \sc{XXV}$ and $\mathrm{Fe} \sc{XXVI}$ lines.

\begin{figure}
    \centering
    \includegraphics[width=0.8\columnwidth]{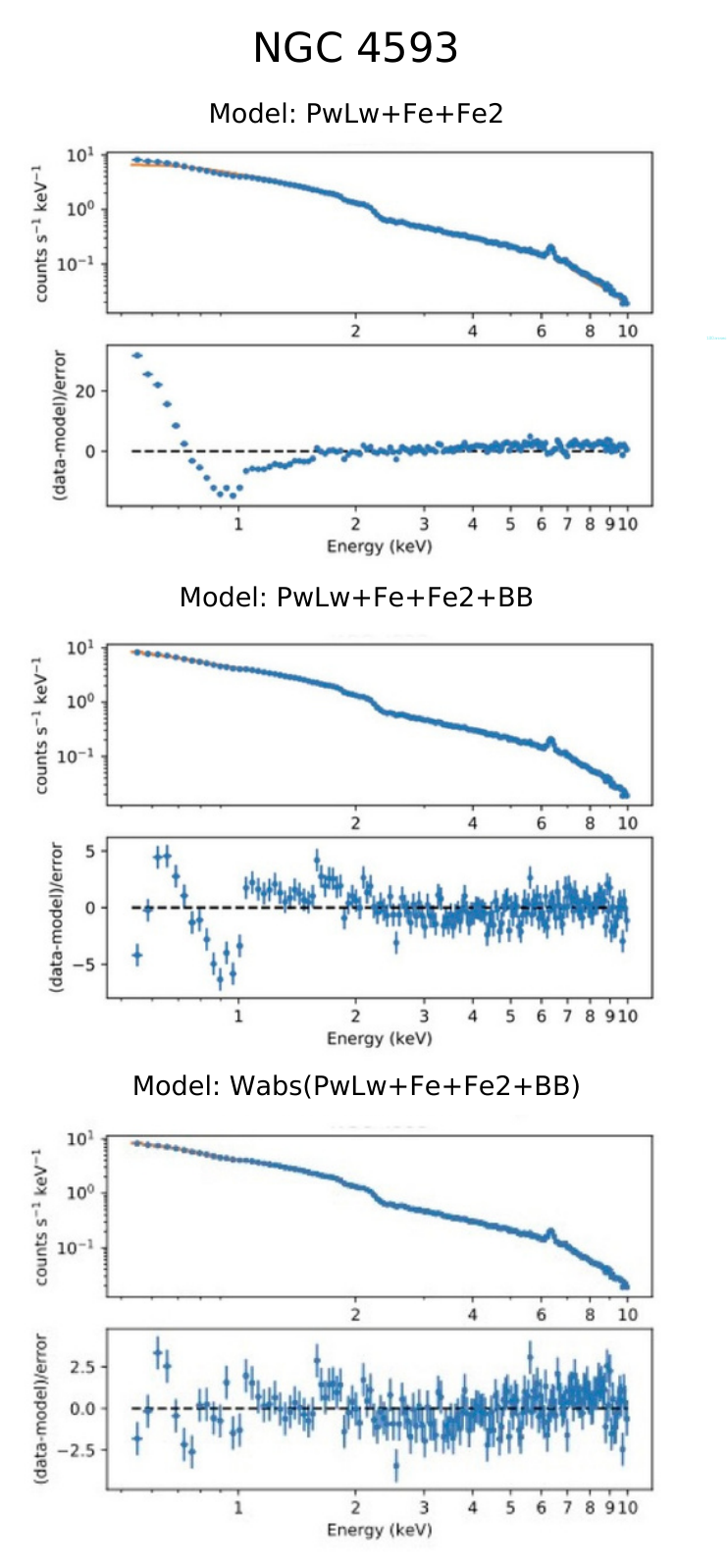}
      \caption{{\it XMM-Newton} {\it EPIC-pn} spectrum of NGC 4593 fitted in the $0.5-10\ \mathrm{keV}$ range. Top panel shows the fit of the baseline model reported in Table \ref{tab:hardband}, middle panel depicts the fit with soft excess added, as shown in Table \ref{tab:se}. Last panel corresponds to the model resulting from the absorption test, a model with warm absorption, see Table \ref{tab:abstest}.}
      \label{fig:N45}
\end{figure}

Extending to the full energy range, we find that soft excess is ubiquitous, see Table \ref{tab:se}, and highly significant in all sources (F-test $> 99\%$). For the black body component, we find temperatures of $\mathrm{k}T<1\ \mathrm{keV}$ for 9 out of the 11 sources. Fairall 51 and ESO 323-G077 yield higher temperatures, however, these values decrease once we add absorption components.

Concerning the absorption test, according to the AIC, we determine the presence of absorption in all the {\it PL-pol} sources, see Table \ref{tab:abstest}. Mrk 1218 and Mrk 231 favour the cold absorption scenario, resulting in column densities on the order of $10^{21}\ \mathrm{cm^{-2}}$. For NGC 4593, IRAS 15091-2107, Mrk 766 and ESO 323-G077, the AIC indicates that warm absorption is the best-fitting scenario. Figure \ref{fig:N45} depicts the fits of NGC 4593 as an example of the baseline model, the addition of soft excess and the absorption test, in which case the favoured scenario corresponds to the warm absorption; see Appendix \ref{App:spec} for the fits of the entire sample. The fit is further improved by adding a second warm absorption component for the cases of Fairall 51, Mrk 704, Was 45, and NGC 3227. For the warm absorbers, all the column densities are within the order of $N_H \sim 10^{20}-10^{22}\ \mathrm{cm^{-2}}$. We obtain a large range of values in the ionization parameters, from $-1.4<\log \xi<5.9$. In particular, for Mrk 1239 the results are inconclusive, with the AIC value indicating that the fit improves by adding an absorption component, but none of the two scenarios (cold vs warm) can be statistically favoured.

\subsubsection{Equatorial-polarized sources}

In the hard band of the {\it EQ-pol} sources,  one exhibits a flat spectrum with a power law index of $\sim 1.1$, while the remaining 7 sources have a photon index in the range of $1.5<\Gamma<2.3$. The $\mathrm{Fe\ K\alpha}$ line is significantly detected in 4 sources: NGC 3783, Mrk 841, Mrk 509 and Akn 120. NGC 3783 also shows the $\mathrm{Fe} \sc{XXVI}$ line, and Akn 120 shows both the $\mathrm{Fe} \sc{XXV}$ and $\mathrm{Fe} \sc{XXVI}$ lines.

In the full energy range, we again obtain a result that supports that the soft excess is ubiquitous, with F-tests $>99\%$ for all the sources. The black body temperatures are all $\mathrm{k}T<1\ \mathrm{keV}$.

As for  the absorption test, unlike the {\it PL-pol} sample, we do not find absorption in all the sources. In 6 out of 8 sources we can determine the presence of absorption. Mrk 876 and Mrk 509 are better fitted without any absorber, according to the small value of the AIC test. Mrk 304 favours the cold absorber scenario, with column density of the order of $N_\mathrm{H} \sim10^{21}\ \mathrm{cm^{-2}}$. I Zw1, Mrk 841, KUV 18217+6419 and Akn 120 favour the scenario of warm absorption, yielding column densities of the orders of $N_\mathrm{H} \sim 10^{20}-10^{21}\ \mathrm{cm^{-2}}$ and a range of ionization parameters of $-0.2<\log\xi<3.0$. None of the fits improved by the addition of a second warm absorption component. NGC 3783, similar to the case of Mrk 1239, results in an AIC that indicates fit improvement by the addition of an absorption component, however, the test cannot conclusively indicate whether the favourable scenario consists in cold or warm absorption.

\begin{table}
\begin{adjustbox}{width=\textwidth}
\begin{tabular}{llllllllr}
\hline
$^{(1)}$ & $^{(2)}$ & $^{(3)}$ & $^{(4)}$ & $^{(5)}$ & $^{(6)}$ & $^{(7)}$ & $^{(8)}$ & $^{(9)}$ \\
\hline
\small{Galaxy} &  \small{Model} &  \small{$N_\mathrm{H}$} & \small{Photon} & \small{Norm} & \small{Fe Norm} & \small{$\mathrm{Fe_{XXV}}$ Norm} & \small{$\mathrm{Fe_{XXVI}}$ Norm} & \small{$\chi^2_{\nu}$} \\
& & $\times 10^{22}$ & \small{Index} & $\times 10^{-3}$ & $\times 10^{-5}$ & $\times 10^{-5}$ & $\times 10^{-5}$ & \\
& & $\mathrm{cm^{-2}}$ & $\Gamma$ & $\mathrm{photons/keV/cm^2/s}$ & $\mathrm{photons/cm^2/s}$ & $\mathrm{photons/cm^2/s}$ & $\mathrm{photons/cm^2/s}$ & \\
\hline\\
\textbf{Polar Polarization}\\
\\
Mrk 1218 & \textbf{PwLw} & - & 1.37 $\pm0.10$ & $0.4_{-0.3}^{+0.3}$ & - & - & - & \bf{0.94}\\
\\
IRAS 15091-2107 & \bf{PwLw+Fe} & - & 1.72 $\pm0.06$ & 2.3 $\pm0.2$ & 1.3 $\pm 0.6$ & - & - & \bf{1.10} \\
\\
NGC 4593 & \bf{PwLw+Fe+Fe2} & - & 1.74 $^{+0.04}_{-0.07}$ & 4.64 $^{+0.2}_{-0.14}$ & 2.8 $\pm0.2$ & - & 5.5 $^{+4.7}_{-4.3}$ & \bf{1.12} \\
\\
Mrk 231 & \bf{N$_H$(PwLw)} & 2.4 $^{+1.8}_{-1.7}$ & 1.0 $\pm0.3$ & 0.006 $^{+0.004}_{-0.002}$ & - & - & - & \bf{1.13}\\
\\  
Fairall 51 & \bf{N$_H$(PwLw+Fe)} & 2.3 $\pm0.2$ & 1.89 $\pm0.04$ & 8.8 $\pm0.7$ & 2.6 $\pm0.5$ & - & - & \bf{1.33}\\
\\
Mrk 704 & \bf{PwLw+Fe} & - & 1.82 $\pm0.02$ & 2.69 $\pm0.06$ & 1.1 $\pm0.2$ & - & - & \bf{1.37}\\
\\ 
NGC 3227 & \bf{N$_H$(PwLw+Fe)} & 0.29 $\pm0.09$ & 1.58 $\pm0.02$ & 0.0062 $\pm0.0002$ & 3.2 $\pm0.3$ & - & - & \bf{1.49}\\
\\
Was 45 & \bf{N$_H$(PwLw+Fe)} & 7.2$^{+0.9}_{-1.0}$ & 1.56 $\pm0.13$ & 0.58 $^{+0.16}_{-0.12}$  & 1.0 $\pm0.2$ & - & - & \bf{1.66}\\
\\
Mrk 766 & PwLw+Fe+Fe2+Fe3 & - & 2.11 & 8.0 & 1.0 & 0.9 & 0.6 & 2.10 \\
\\
ESO 323-G077 & N$_H$(PwLw+Fe+Fe2) & 5.5 & 0.6 & 0.11 & 2.0 & - & 0.9 & 5.54\\
\\
Mrk 1239 & N$_H$(PwLw+Fe) & 57.7 & 4.3 & 186 & 1.6 & - & - & 6.0\\
\\
\hline \\
\textbf{Equatorial Polarization}\\
\\
Mrk 876 & \textbf{PwLw} & - & 1.62 $0.11$ & 1.7 $\pm0.3$ & - & - & - & \textbf{0.86} \\
\\
NGC 3783 & \textbf{N$_H$(PwLw+Fe+Fe2)} & 0.6 $\pm0.2$ & 1.50 $\pm0.03$ & 4.9 $_{-0.2}^{+0.3}$ & 4.5 $\pm0.4$ & - & 1.3 $\pm0.4$ & \textbf{1.65} \\
\\
Mrk 841 & \textbf{N$_H$(PwLw+Fe)} & 0.7 $\pm0.4$ & 1.52 $\pm0.06$ & 1.5 $\pm0.2$ & 1.0 $\pm0.2$ & - & - & \textbf{1.75} \\
\\
I Zw1 & \textbf{PwLw} & - & 2.33 $\pm0.05$ & 2.7 $\pm0.2$ & - & - & - & \textbf{1.85}\\
\\ 
Mrk 509 & PwLw+Fe & - & 1.86 & 8.4 & 1.7 & - & - & 2.30\\
\\ 
Akn 120 & PwLw+Fe+Fe2+Fe3 & - & 1.93 & 9.9 & 2.2 & 0.7 & 0.8 & 2.50 \\
\\
KUV 18217+6419 & PwLw & - & 1.14 & 5.8 & - & - & - & 2.53 \\
\\ 
Mrk 304 & N$_H$*PwLw & 5.99 & 1.74 & 1.14 & - & - & - & 3.0\\
\\
\hline
\end{tabular}
\end{adjustbox}
\caption{{\bf Hard band fits.} Best fit parameter values and goodness of fit for the hard band analysis. Column 2 indicates the resulting model. In columns 3 to 8, we report the value of the model parameters: column density, (if significant), photon index and power law normalization, and normalization for the significant Fe emission lines. The resulting model statistics is shown in column 9. Values in bold correspond to fits with $\chi^2_\nu<2.0$ for which we can calculate parameter errors.}
\label{tab:hardband}
\end{table}

\begin{table}
\begin{adjustbox}{width=\textwidth}
\begin{tabular}{llllllrc}
\hline
$^{(1)}$ & $^{(2)}$ & $^{(3)}$ & $^{(4)}$ & $^{(5)}$ & $^{(6)}$ & $^{(7)}$ & $^{(8)}$ \\
\hline
\small{Galaxy} & \small{Model} & \small{Photon} & \small{Norm} & \small{$\mathrm{k}T$} & \small{BBnorm} & \small{$\chi^2_{\nu}$} & \small{F-test}\\
& & \small{Index} & $\times 10^{-3}$ & & $\times 10^{-4}$ & & \\
& & $\Gamma$ & $\mathrm{photons/keV/cm^2/s}$ & $\mathrm{keV}$ & $L_{39}/[D_{10}(1+z)]^2$ & & \\
\hline\\
\textbf{Polar Polarization}\\ \\
Mrk 1218 & PwLw & 0.95 & 0.2 & - & - & 2.24 & \\
& PwLw+BB & 0.84$^{+0.10}_{-0.11}$ & 0.13$\pm0.02$ & 0.84 $^{+0.11}_{-0.09}$ & 0.1$\pm0.2$ & \textbf{1.18} & 99.99\%\\ \\
Mrk 231 & PwLw & 0.91 & 0.04 & - & - & 4.30 & \\
& PwLw+BB & 0.46 $\pm0.08$ & 0.023 $\pm$ 0.003 & 0.169 $_{-0.012}^{+0.014}$ & 0.0116 $\pm$ 0.0012 & \textbf{1.68} & 100\%\\ \\
IRAS 15091-2107 & PwLw+Fe & 1.44 & 1.4 & - & - & 6.47 &\\ 
& PwLw+Fe+BB & 1.31$\pm0.03$ & 0.93$\pm0.04$ & 0.52$\pm0.02$ & 0.24$\pm0.02$ & \textbf{1.98} & 100\%\\ \\
Mrk 704 & PwLw+Fe & 1.99 & 3.1 & - & - & 25.81 & \\
& PwLw+Fe+BB & 1.83 & 3.0 & 0.07 & 2.0 & 2.29 & 100\% \\ \\
NGC 4593 & PwLw+Fe+Fe2 & 1.78 & 5.0 & - & - & 25.32 &\\
& PwLw+Fe+Fe2+BB & 1.69 & 4.0 & 0.07 & 1.3 & 2.60 & 100\% \\ \\
Was 45 & PwLw+Fe & 0.23 & 0.04 & - & - & 9.24 &\\
& PwLw+Fe+BB & 0.14 & 0.04 & 0.09 & 0.03 & 6.9 & 99.99\%\\ \\
ESO 323-G077 & PwLw+Fe+Fe2 & 0.3 & 0.05 & - & - & 32.5 &\\
& PwLw+Fe+Fe2+BB & 0.3 & 0.07 & 3.4 & 0.7 & 8.10 & 100\%\\ \\
Mrk 766 & PwLw+Fe+Fe2+Fe3 & 2.25 & 8.3 & - & - & 92.02 &\\
& PwLw+Fe+Fe2+Fe3+BB & 2.06 & 7.3 & 0.08 & 3.2 & 13.45 & 100\%\\ \\
NGC 3227 & PwLw+Fe & 1.37 & 5.0 & - & - & 40.28 & \\
& PwLw+Fe+BB & 1.41 & 4.0 & 0.87 & 0.49 & 27.53 & 99.99\% \\ \\
Fairall 51 & PwLw+Fe & 1.06 & 2.0 & - & - & 56.97 &\\
& PwLw+Fe+BB & 1.24 & 14.0 & 1.2 & 1.5 & 37.57 & 99.99\%\\ \\
 Mrk 1239 & PwLw+Fe & 3.38 & 0.14 & - & - & 70.38 & \\
 & PwLw+Fe+BB & -0.14 & 0.013 & 0.06 & 0.15 & 31.41 & 100\%\\ \\
\hline\\
\textbf{Equatorial Polarization}\\ \\
Mrk 876 & PwLw & 1.99 & 2.5 & - & - & 2.55 & \\ 
& PwLw+BB & $1.73 \pm 0.03$ & $1.64 \pm 0.06$ & $0.102 \pm 0.006$ & $0.48_{-0.05}^{+0.6}$ & \textbf{1.09} & 100\% \\ \\
I Zw1 & PwLw & 2.6 & 3.2 & - & - & 4.50 & \\
& PwLw+BB & 2.3 & 2.7 & 0.11 & 0.4 & 2.25 & 100\% \\ \\      
Mrk 841 & PwLw+Fe & 1.81 & 1.8 & - & - & 24.81 & \\ 
& PwLw+Fe+BB & 1.46 & 1.3 & 0.101 & 0.6 & 2.36 & 100\% \\ \\
KUV 18217+6419 & PwLw & 1.55 & 10.1 & - & - & 18.50 & \\
& PwLw+BB & 1.15 & 5.9 & 0.2 & 1.7 & 2.63 & 100\%\\ \\
Akn 120 & PwLw+Fe+Fe2+Fe3 & 2.11 & 12.0 & - & - & 21.43 & \\ 
& PwLw+Fe+Fe2+Fe3+BB & 1.97 & 10.5 & 0.14 & 0.8 & 4.20 & 100\% \\ \\
Mrk 509 & PwLw+Fe & 2.11 & 11.0 & - & - & 33.20 & \\
& PwLw+Fe+BB & 1.96 & 9.6 & 0.102 & 1.4 & 4.75 & 100\% \\ \\ 
Mrk 304 & N$_H$*PwLw & 0.54 & 0.11 & - & - & 12.16 &\\ 
& N$_H$*PwLw+BB & 0.46 & 0.10 & 0.07 & 0.2 & 8.43 & 99.99\%\\ \\
NGC 3783 & PwLw+Fe+Fe2 & 1.28 & 3.4 & - & - & 40.38 & \\ 
& PwLw+Fe+Fe2+BB & 1.20 & 3.0 & 0.07 & 2.8 & 14.80 & 100\%\\ \\
 \hline\\
\end{tabular}
\end{adjustbox}
\caption{{\bf Full energy band.} For each source, the first row shows the baseline model, the second row corresponds to the model with black body as soft excess. Columns 3 and 4 report the power law index and normalization. Columns 5 and 6 show the black body temperature and normalization. The following columns, 7 and 8, correspond to the fit statistics. Last column indicates the F-test resulting from comparing models with and without soft excess. We highlight in bold the fits that yield $\chi^2_\nu<2$.}
\label{tab:se}
\end{table}

\begin{table}
\begin{adjustbox}{width=\textwidth}
\begin{tabular}{llllllllcrrrc}
\hline
$^{(1)}$ & $^{(2)}$ & $^{(3)}$ & $^{(4)}$ & $^{(5)}$ & $^{(6)}$ & $^{(7)}$ & $^{(8)}$ & $^{(9)}$ & $^{(10)}$ & $^{(11)}$ & $^{(12)}$ & $^{(13)}$\\
\hline
\small{Galaxy} & \small{Model} & \small{nH} & \small{$\log \xi$} & \small{$\mathrm{N_{H2}}$} & \small{$\log \xi_2$} & \small{Photon} & \small{$\mathrm{k}T$} & \small{Lum$\mathrm{_{X-Ray}}$} & \small{$\chi^2_{\nu}$} & \small{$AIC_{\mathrm{BB}}$} & \small{AIC} &\small{$1/F_{\mathrm{AIC}}$} \\
& & \small{$\times 10^{22}\mathrm{cm^{-2}}$} & & \small{$\times 10^{22}\mathrm{cm^{-2}}$} & & \small{Index $\Gamma$} & $\mathrm{keV}$ & \small{$\times 10^{43}\mathrm{erg\ s^{-1}}$} & & & &\\
\hline \\
\textbf{Polar Polarization}\\ \\
Mrk 1218 & \textbf{ColdAbs} & 0.47$^{+0.14}_{-0.12}$ & - & - & - & 1.37$\pm0.09$ & 0.062$\pm0.012$ & 0.604 & \textbf{1.17} & 151.81 & 113.66 & $1.9 \times 10^{8}$ \\
 & WarmAbs & 0.15 $\pm$ 0.10 & 1.2 $_{-1.6}^{+0.3}$ & - & - & 1.0 $^{+1.0}_{-0.3}$ & 0.80 $_{-0.10}^{+0.80}$ & & 1.21 & & 118.18 & $2.01 \times 10^{7}$\\ \\
Mrk 231 & \textbf{ColdAbs} & 0.6 $\pm$ 0.2 & - & - & - & 0.65 $\pm$ 0.10 & 0.106 $_{-0.012}^{+0.015}$ & 0.324 & \textbf{1.29} & 140.41 & 110.78 & $2.7\times10^{6}$\\
& WarmAbs & 0.05 & 0.4 & - & - & 0.48 $_{-0.07}^{+0.10}$ & 0.168 $\pm0.013$ & & 1.82 & & 152.41 & 0.002\\ \\
NGC 4593 & ColdAbs & 0.060 $\pm0.010$ & - & - & - & 1.735 $\pm0.009$ & 0.0782 $^{+0.0013}_{-0.0014}$ & & 1.94 & 449.37 & 334.59 & $8.4\times10^{24}$\\
& \textbf{WarmAbs} & 0.14$\pm0.02$ & 2.36$\pm0.08$ & - & - & 1.727 $\pm0.007$ & 0.079 $\pm0.003$ & 0.437 & \textbf{1.40} & & 253.78 & $3.0\times10^{42}$\\ \\
Fairall 51 & ColdAbs & 0.0 & - & - & - & 1.24 & 1.15 & & 37.80 & 6095.58 & 6097.58 & 0.37\\
& WarmAbs & 1.31 & 0.57 & - & - & 1.81 & 0.11 & & 3.35 & & 549.52 & $>10^{90}$\\
 & \textbf{2WarmAbs} & 2.50 $^{+1.03}_{-0.84}$ & 2.5 $^{+0.5}_{-0.2}$ & 1.0 $^{+0.4}_{-0.2}$ & 0.46 $^{+0.05}_{-0.04}$ & 1.86 $\pm 0.09$ & 0.113 $^{+0.006}_{-0.004}$ & 1.268 & \textbf{1.60} & & 269.84 & $>10^{90}$\\ \\
IRAS 15091-2107 & ColdAbs & 0.08 $\pm$ 0.03 & - & - & - & 1.51 $_{-0.07}^{+0.06}$ & 0.51 $_{-0.03}^{+0.04}$ & & 1.83 & 301.57 & 278.51 & $1.02\times10^{5}$\\ 
& \textbf{WarmAbs} & 0.12 $_{-0.12}^{+0.05}$ & -0.46 $_{-0.23}^{+0.14}$ & - & - & 1.63 $_{-0.07}^{+0.10}$ & 0.45 $_{-0.11}^{+0.07}$ & 4.562 & \textbf{1.63} & & 251.32 & $8.2\times10^{10}$\\ \\
Mrk 704 & ColdAbs & 0.05 & - & - & - & 1.88 & 0.07 & & 2.05 & 382.46 & 344.67 & $1.6\times10^{8}$\\
 & WarmAbs & 0.13 $_{-0.04}^{+0.03}$ & 2.15 $_{-0.15}^{+0.14}$ & - & - & 1.882 $_{-0.013}^{+0.012}$ & 0.076 $\pm$ 0.003 & & 1.84 & & 318.36 & $8.3\times10^{13}$\\
 & \textbf{2WarmAbs} & 0.061 $^{+0.002}_{-0.009}$ & 0.28 $^{+0.46}_{-0.13}$ & 0.23 $^{+0.02}_{-0.14}$ & 2.44 $^{+0.60}_{-0.11}$ & 1.9 $\pm0.9$ & 0.113 $^{+0.113}_{-0.003}$ & 2.919 & \textbf{1.73} & & 293.35 & $2.2\times10^{19}$\\ \\
 Was 45 & ColdAbs & 1.4 & - & - & - & 0.6 & 0.05 & & 4.54 & 858.41 & 565.55 & $>10^{90}$\\
& WarmAbs & 2.74 & 0.28 & - & - & 1.17 & 0.09 & & 2.98 & & 374.06 & $2.0\times10^{62}$\\
& \textbf{2WarmAbs} & 0.013 & 2.04 & 2.5 & 0.3 & 1.30 & 0.09 & 0.325 & 3.67 & & 454.79 & $4.4\times10^{87}$\\ \\
Mrk 766 & ColdAbs & 0.09 & - & - & - & 2.13 & 0.08 & & 11.72 & 2220.43 & 1926.75 & $5.9\times10^{63}$\\
& \textbf{WarmAbs} & 0.4 & 0.9 & - & - & 2.18 & 0.12 & 0.128 & 2.87 & & 480.05 & $>10^{90}$\\ \\
NGC 3227 & ColdAbs & 0.0 & - & - & - & 1.4 & 0.9 & & 27.71 & 4635.78 & 4640.26 & 0.11\\
& WarmAbs & 0.4 & 1.6 & - & - & 1.51 & 0.63 & & 6.12 & & 1023.29 & $>10^{90}$\\
& \textbf{2WarmAbs} & 0.6 & 0.3 & 0.02 & 5.9 & 1.59 & 0.07 & 0.144 & 3.72 & & 627.24 & $>10^{90}$\\ \\ 
ESO 323-G077 & ColdAbs & 0.97 & - & - & - & 0.58 & 0.09 & & 15.84 & 1259.63 & 2436.98 & X\\
& \textbf{WarmAbs} & 0.3 & -1.4 & - & - & 0.14 & 0.12 & 0.230 & 6.64 & & 1027.53 & $2.5\times10^{50}$\\ \\
Mrk 1239 (*) & ColdAbs & 0.02 & - & - & - & -0.15 & 0.15 & & 31.06 & 10495.39 & 4390.27 & $>10^{90}$\\
& WarmAbs & 0.05 & 5.96 & - & - & -0.16 & 0.15 & 0.159 & 31.45 & & 4415.32 & $>10^{90}$\\ \\
\hline \\
\textbf{Equatorial Polarization}\\ \\
Mrk 876 & \textbf{Unabs} & - & - & - & - & $1.73 \pm 0.03$ & $0.102 \pm 0.006$ & 0.341 & \textbf{1.09} & 106.95 & - & -\\ 
& ColdAbs & 0.0 & - & - & - & 1.73$^{+0.06}_{-0.05}$ & 0.110$^{+0.014}_{-0.015}$ & & 1.10 &  & 108.96 & 0.37\\
& WarmAbs & 0.04$\pm0.04$ & 0.2$^{+0.2}_{-1.4}$ & - & - & 1.73$^{+0.06}_{-0.05}$ & 0.13$^{+0.02}_{-0.03}$ & & 1.19 & & 117.70 & 0.005\\ \\
1Zw 1 & ColdAbs & 0.0 & - & - & - & 2.80$\pm0.03$ & 1.0$^{+0.004}_{-1.0}$ & & 3.02 & 307.83 & 409.14 & X \\
& \textbf{WarmAbs} & 0.09$^{+0.04}_{-0.09}$ & -0.2$^{+0.6}_{-0.2}$ & - & - & 2.3$^{+0.3}_{-0.4}$ & 0.12$^{+0.011}_{-0.005}$ & 8.648 & \textbf{1.91} & & 261.57 & $1.11\times10^{10}$\\ \\
Mrk 841 & ColdAbs & 0.0 & - & - & - & 1.45 & 0.102 & & 2.38 & 375.48 & 378.51 & 0.22\\ 
& \textbf{WarmAbs} & 0.09 & 1.04 & - & - & 1.46 & 0.12 & 3.349 & 2.15 & & 342.26 & $1.6\times10^{7}$\\ \\
KUV 18217+6419 & ColdAbs & 0.0 & - & - & - & 1.15 & 0.2 & & 2.65 & 445.24 & 447.61 & 0.3\\
& \textbf{WarmAbs} & 1.0 & 3.28 & - & - & 1.15 & 0.2 & 1370.2 & 2.47 & & 417.12 & $1.3\times10^{6}$\\ \\
Akn 120 & ColdAbs & 0.0 & - & - & - & 1.98 & 0.14 & & 3.66 & 617.45 & 619.45 & 0.4\\
& \textbf{WarmAbs} & 0.30 & 0.27 & - & - & 1.98 & 0.15 & 11.419 & 3.49 & & 587.54 & $3.1\times10^{6}$ \\ \\
Mrk 509 & \textbf{Unabs} & - & - & - &- & 1.96 & 0.102 & 12.454 &4.75 & 802.98 & - & - \\
& ColdAbs & 0.0 & - & - & - & 1.95 & 0.103 & & 4.78 & & 805.77 & 0.3\\
& WarmAbs & 0.014 & 4.6 & - & - & 1.99 & 0.11 & & 6.29 & & 1052.53 & X\\ \\
Mrk 304 & \textbf{ColdAbs} & 0.96 & - & - & - & 0.85 & 0.05 & 3.335 & 6.03 & 1146.04 & 817.48 & $2.2\times10^{71}$\\
& WarmAbs & 0.05 & 0.02 & - & - & 0.51 & 0.08 & & 7.92 & & 1064.92 & $4.1\times10^{17}$\\ \\
NGC 3783 (*) & ColdAbs & 0.38 & - & - & - & 1.45 & 0.07 & & 2.82 & 2453.57 & 477.19 & $>10^{90}$\\
& WarmAbs & 0.25 & -0.43 & - & - & 1.42 & 0.09 & 0.688 & 7.13 & & 1178.32 & $>10^{90}$ \\ \\
 \hline
\end{tabular}
\end{adjustbox}
\caption{{\bf Absorption test}. The model in bold corresponds to preferred model. We indicate (Unabs) for unabsorbed sources and use (*) for cases where $1/F_{\mathrm{AIC}}$ does not favour one particular model. In columns 3 to 8, we report the corresponding model parameters in the following order: column densities and ionization parameter in the case of warm absorbers, power law index and black body temperature. In column 9 we report the X-ray luminosity of the resulting model\footnote{Luminosities for Mrk 1239 and NGC 3783 are calculated with the warm absorption model, as is the one reported in literature.}. Column 10 indicate the fit statistics, columns 11 and 12 correspond to the AIC, where AIC$_{\mathrm{BB}}$ corresponds to the model with absorption. Column 13 indicates the factor $1/F_{\mathrm{AIC}}$, an indicator of the fit improvement.}
\label{tab:abstest}
\end{table}

\section{Discussion} \label{sect:Disc}

\begin{figure}
 \centering
    \includegraphics[scale=0.27]{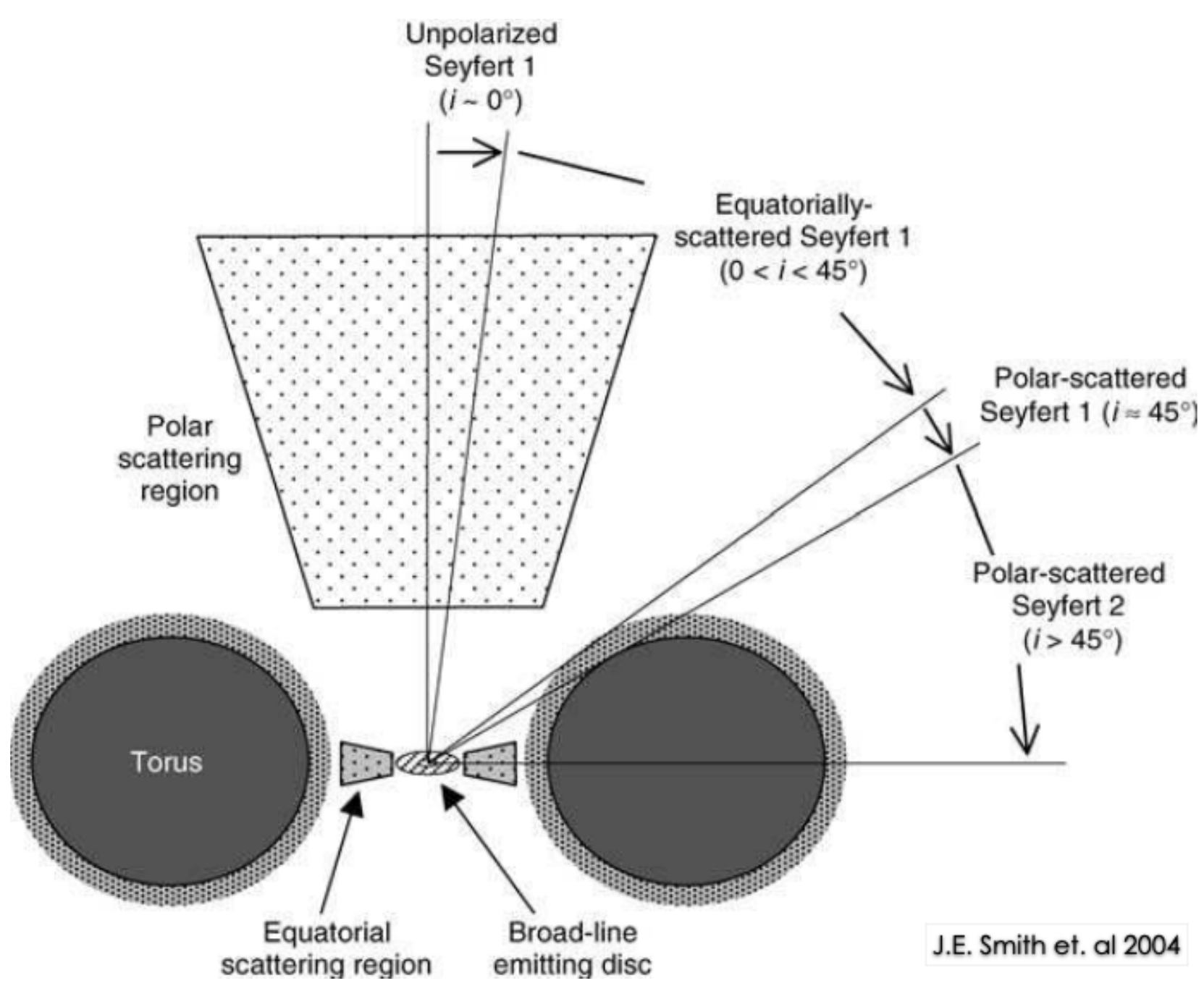}
    \caption{The unification scheme where both scattering regions are present in all Seyfert galaxies. According to this scheme, the observed optical polarization is due to the orientation of the AGN towards our line-of-sight, \citep{Smith2004}.}
    \label{fig:Smith}
\end{figure}

Our analysis on 19 Sy sources is intended to test the presence of absorption in the X-ray spectra in the context of the unification scheme based on optical polarization. Figure \ref{fig:Smith} shows the schematic view where we see both bi-conical and co-planar scattering regions and the line-of-sight orientation corresponding to the different observed scenarios.

We begin discussing the common X-ray properties of our Sy 1 sample, i.e. our baseline model, {\it PowerLaw + Fe + BB}. The main continuum component is well fitted in the hard band for the majority of the sample. A photon index in the range of $1.4<\Gamma<2.3$ is considered to be within the typical range for type 1 Sy galaxies, e.g. \citep{Pi2005, C2006, Si2011, Co2011, Rui2017}, and so is the case for 15 out of our 19 sources. Also consistently with studies on X-ray properties of Sy 1, we find the $\mathrm{Fe\ K\alpha}$ line in 13 out of the 19 observations, corresponding to the narrow component of the line, associated to the molecular torus, \citep{Ri20142, Ri2014}. We did not consider any broadening effect on the line, which would place the line emission mechanism closer to the accretion disk. Extending to the full energy band, we find the soft excess to be ubiquitous, resulting in an average temperature of $0.126\ \mathrm{keV}$. While our sample can be fitted by considering these common characteristics of type 1 Sy, it is worth mentioning that a Seyfert Galaxy spectrum can be affected by additional complex processes, whose detailed description is beyond the scope of our analysis. The effects of Compton reflection, for example, might affect the shape of the continuum making it different from a simple power law. In the case of many sources in our sample, the spectra prove to be more complex, possibly requiring additional components. For this reason, we report in Section \ref{sect:indiv}, results of more detailed spectral analyses from the literature for each individual source. When available, we cite X-ray analysis carried out on the same XMM-Newton observation as the one used in our work.

Regarding the absorption test, we first point out that the characteristics here reported on the warm absorbers, column density and ionization parameter, are consistent with the description provided by \citet{L2021}: the column densities range is $N_\mathrm{H} \sim 10^{20}-10^{22}\ \mathrm{cm^{2}}$, and the ionization parameters range is $-1.0<\log \xi<3.0$. Only NGC 3227 yields a more highly ionized warm absorber, $\log\xi\sim6.0$, a result that could be further improved by a more detailed modeling, (see notes on NGC 3227 in section \ref{sect:indiv}). The sources that favor the cold absorption scenario yield column densities $N_\mathrm{H}\sim 10^{21}\ \mathrm{cm^{2}}$, consistent with type 1 sources that are not heavily absorbed. In general, the detected column density is always consistent with that of type 1 AGN, with the dividing value between type 1 and type 2 sources being $N_H \sim 10^{22}\ \mathrm{ cm^{-2}}$, \citep{Volk2012}.

Concerning the presence of absorption, our analysis shows that the absorption is present in $100\%$ of the {\it PL-pol} sources. In particular $\sim 73\%$ (8/11) of these sources favor the warm absorption scenario and only $18\%$ (2/11) sources are affected by a cold absorber. The presence of warm absorption in the {\it PL-pol} sources can be put in the context of the scenario described by \citet{Smith2004} by interpreting it as the presence of the outer layer of the torus in the line-of-sight, which gets ionized by the central engine. \citet{Bl2005} argues that warm absorbers in Seyfert galaxies are more likely to originate in outflows from dusty torus, lending support to the hypothesis that the warm absorber gas could be located in its the outer layers. The two {\it PL-pol} sources that resulted affected by cold absorption yield low column densities, $N_{\mathrm{H}}\sim 10^{21}\ \mathrm{cm^{-2}}$, we can interpret this as a result of our line of sight passing through a colder section of the torus atmosphere but remaining on the type 1 regime. 

In slight contrast, $75\%$ (6/8) of the {\it EQ-pol} sources are affected by absorption, with $50\%$ (4/8) favoring the warm absorption scenario. This fraction is consistent with previous studies on the spectra of type 1 Seyferts, \citep[e.g.][]{Rey1995, L2014I}, where they respectively report that $50\%$ up to $65\%$ of Sy 1 show the presence of warm absorbers. For two {\it EQ-pol} sources, $25\%$ of the sample, our analysis does not find a significant absorption component, consistent with the expected description of a type 1 AGN where our line-of-sight is looking directly into the central region. 

In order to determine if the difference percentage of absorbed sources described above is significantly different, we conducted a Kolmogorov-Smirnov test (KS) at $95\%$ confidence. The result of the test yields a p-value of $\sim0.87$. A p-value much greater than the significance level of 0.05 ($95\%$) indicates that we reject the test's null hypothesis, i.e., that the samples are drawn from  the same distribution. This means that the two sub-samples -- {\it PL-pol} and {\it EQ-pol} -- are statistically different from each other in terms of the incidence of absorption in these X-ray data.

\begin{figure}
    \centering
    \includegraphics[scale=0.25]{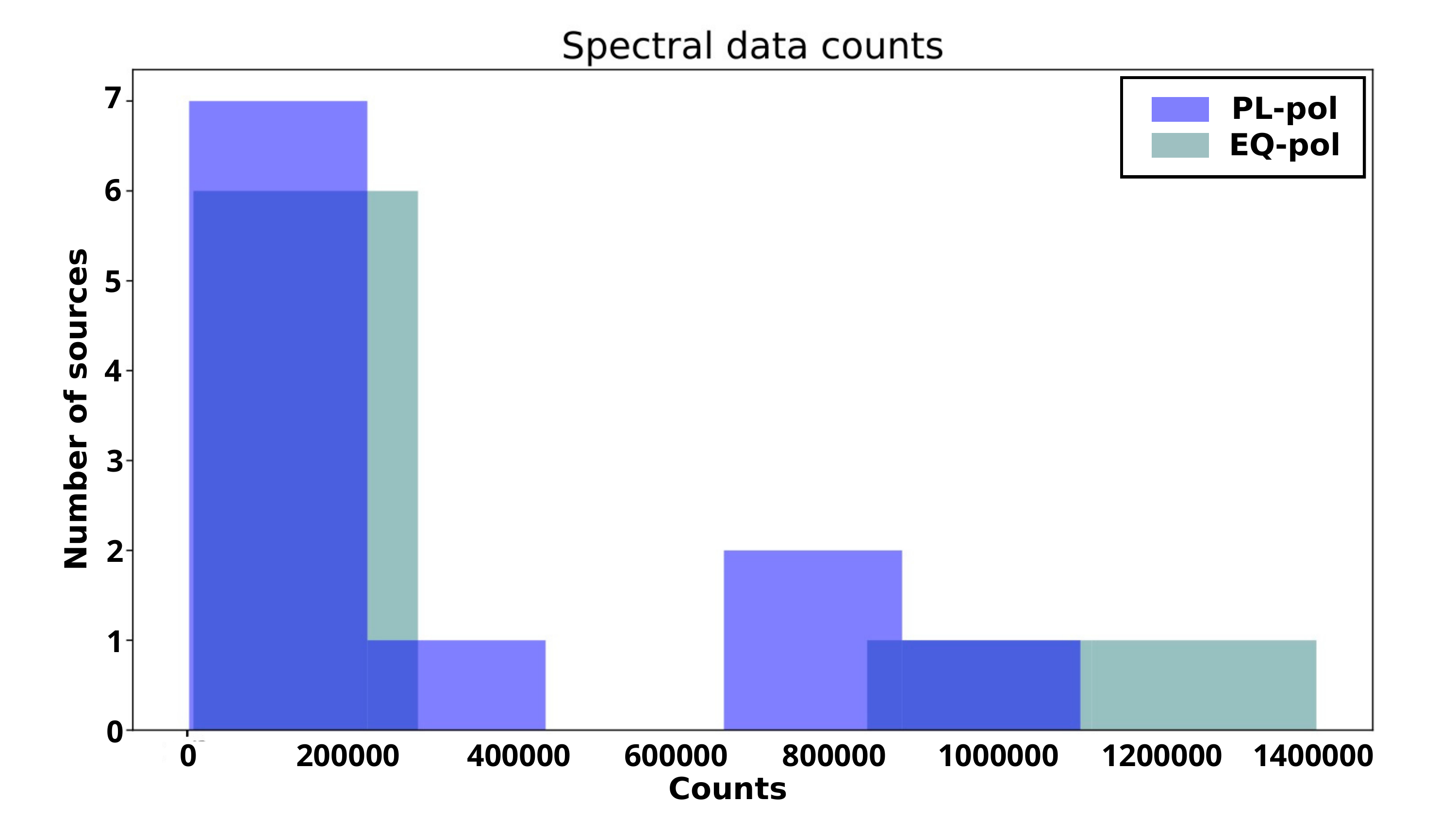}
    \caption{Spectral data counts of each source.}
    \label{fig:counts}
\end{figure}

\begin{figure}
    \centering
    \includegraphics[scale=0.25]{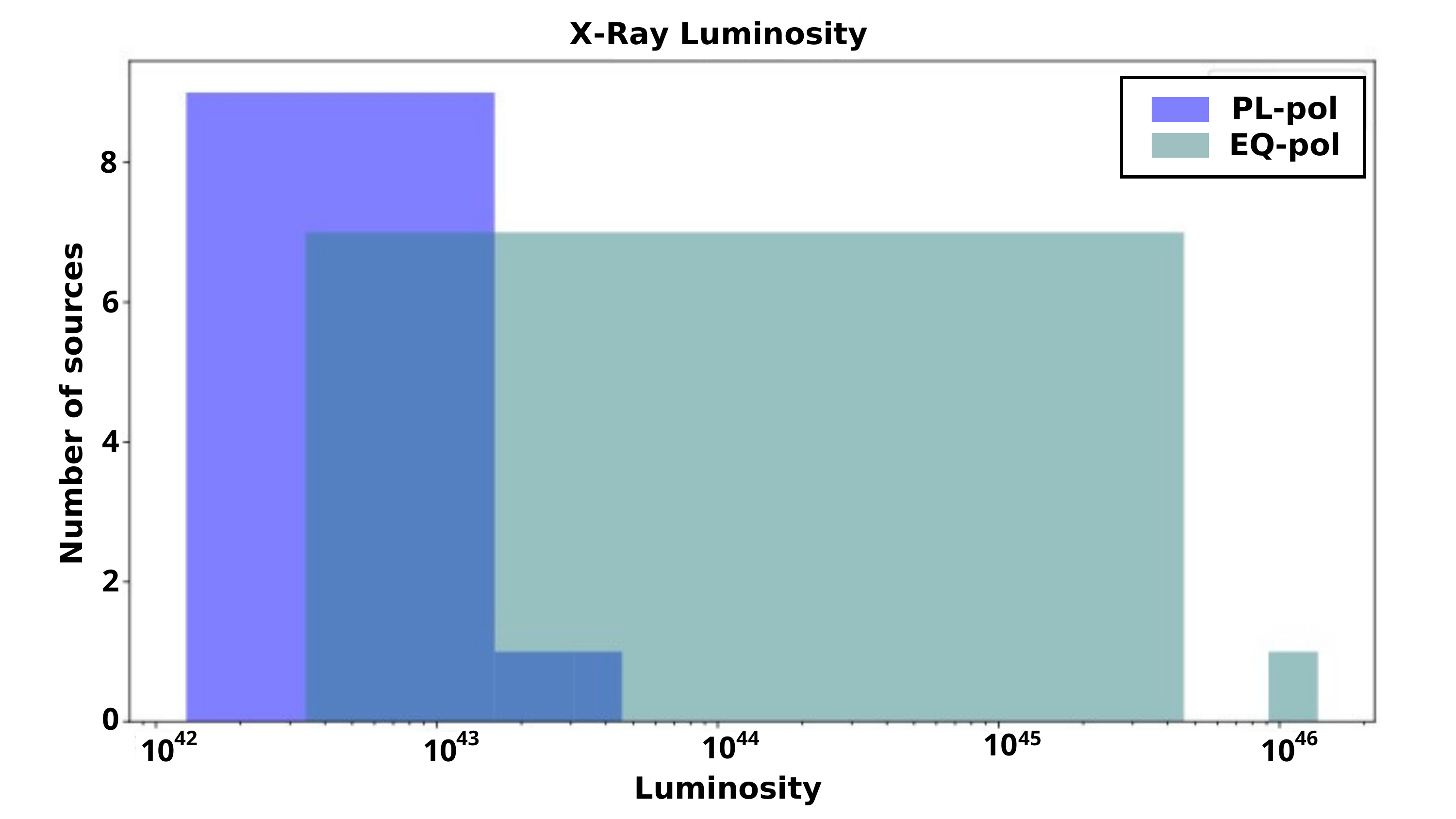}
    \caption{\textbf{X-ray Luminosity ($0.5-10\ \mathrm{keV}$)} of each source. The value is shown in Table \ref{tab:abstest}, referring to the luminosity of the preferred model after the absorption test.}
    \label{fig:lumos}
\end{figure}

We also considered the possible bias due to the data selected for this analysis. Our selection criteria was to choose the longest exposure time for each source, and this choice may introduce a bias related to the signal-to-noise of the X-ray spectra contained in of each of the two sub-samples.  Figure \ref{fig:counts} is an histogram of the number of counts for each source, color-divided by sub-samples. We can see that there is no significant distinction between the two sub-samples concerning the quality of the observation in terms of number of X-ray counts, which, if present, might have hindered the modeling of the spectrum. Regarding the X-ray luminosity of the sources, which may be intuitively associated to a higher frequency of the absorber due to the photoionization process, we show in Figure \ref{fig:lumos} an histogram of the X-ray luminosity calculated for the model resulting from the absorption test. This plot shows that the {\it EQ-pol} sources are actually more luminous than the {\it PL-pol}, therefore discarding the hypothesis that a higher luminosity is directly associated to the ubiquitous presence of an absorber.

A more recent study, \citet{Af2019}, identifies 4 sources as {\it EQ-pol} that are instead listed as {\it PL-pol} in the studies by Smith, S04: Mrk 231, Mrk 704, NGC 3227, and NGC 4593. By assuming this new classification, the fraction of polar and equatorial polarized sources would shift to 7 and 12, respectively. For the former group, with all of them affected by absorption, only one favors the model with cold-absorption and 6, $\sim 86\%$, the warm-absorption scenario. For the sub-sample of {\it EQ-pol} sources, $\sim 83\%$ (10/12) of the sources are affected by absorption, with $\sim 58\%$ (7/12) favoring the presence of at least one warm absorber. This incidence of warm absorption is still within the threshold of the numbers reported by \citep{L2014I}, therefore we concluded that assuming an updated classification does not change the first-order conclusion of this analysis. For this reason and for consistency with the original criterion that defined our sample, we prefer to stick to the original classification proposed by Smith as a first test for the polarization unified scheme.

Now considering the sample of polarized sources as a whole, $\sim 90\%$ (17/19) of our sample shows presence of absorption and we can confirm the presence of at least one warm absorber in $\sim 63\%$ of the polarized Seyfert 1. This in itself is an interesting result on the incidence of X-ray absorption on polarized Seyfert galaxies, especially when compared to previous studies \citep[e.g.][]{Rey1995, L2014I}, where a very similar percentage of X-ray warm absorbers was reported ($50\%$ up to $65\%$, respectively).

From our review on previously published analysis on each source, reported in the following section, we find that the literature modeling is consistent with our analysis as a first approximation. In the majority of the sources, we see that best fit models include more components, e.g. reflection affecting the shape of the continuum as in the cases of Mrk 1239, Mrk 766, ESO 323-G077. More specifically, we highlight that the modeling of absorption often requires more than one absorbing layer. In the majority of the sources, we see absorption modeled by two or even more components. In some cases, partial covering is the best way to reproduce the shape of the spectrum. We remark that the optical and X-ray data analyzed herein are far from being simultaneous since the majority of the polarization data were obtained in 1996-1999, i.e. prior to XMM-Newton launching. Although desirable, obtaining fully simultaneous data of this sample would require considerable observational effort. We are confident that this analysis may serve to lay the basis for future observations.

These considerations motivate a more in-depth analysis addressed to characterize the absorber. For the time being, our analysis can serve as a first approximation model. Here, we describe the common components of a type 1 Seyfert spectrum and establish a point of reference for the presence of X-ray absorption in a sample of polarized Seyferts. This study would also benefit by the addition of a larger sample of Seyfert galaxies with well known and reported polarization.

\subsection{Notes on individual sources}\label{sect:indiv}

In this section, we briefly present properties taken from the available literature on each source of our sample. When possible, we report published results based on the analysis of the same data set used in our work.\\

\subsubsection{Polar-polarized sources}

\textbf{\it{Mrk 1218}}\\

\citet{HG2017} concludes that this source is best fitted by a power law with a cold absorber of $N_{\mathrm{H}}\sim 9\times10^{20}\ \mathrm{cm^{-2}}$ in the soft energy range. This result is consistent with our analysis, (see Table \ref{tab:abstest}). \\

\textbf{\it{Mrk 704}}\\

Our best fit for this source includes 2 warm absorbers. This result is consistent with what is reported in \citet{L2011} \& \citet{Ma2011}, where the absorption is interpreted as the line of sight passing close to the torus. As an example of the complexity that these sources show, \citet{Ma2011} produced a best fit model including a second soft excess element and partial covering by a cold absorber.\\

\textbf{\it{Mrk 1239}}\\

\citet{Bu2020,Bu2023} report a very detailed analysis on this NLSy1. The hard band requires a relativistic blurred reflection component besides the power law which is partially covered by ionized material. This source has starburst activity, which affects and even dominates the soft band. Since we do not account for this features, our model fails to fit the continuum and the absorption test results inconclusive.\\

\textbf{\it{NGC 3227}}\\

In our analysis, the results favour the model with 2 warm absorbers. This result is consistent with the work by \citet{Mar2009}, where the best fit model finds 2 warm absorbers, besides a strong soft excess also absorbed by cold gas. Moreover, by estimating a maximum distance, they place the ionized absorber outside the BLR. \citet{N2021} also finds 2 warm absorbers, as well as a partially covered power law and a reflection component.\\

\textbf{\it{Was 45}}\\

Our results favour the presence of at least 2 warm absorbers. However, there are features in the spectra that are not accounted for by our model. A more detailed analysis will be reported in Gudi\~no et al. in prep.\\

\textbf{\it{Mrk 766}}\\

We find a power law continuum on the steeper end of the typical Sy range, $\Gamma \sim2.2 $, as well as three different Fe emission lines and presence of a warm absorber. This result is consistent with the studies made by \citet{Mi2006} \& \citet{T2007}. In particular, with the reports of a warm absorber with $\log \xi \sim 1$ and $N_{\mathrm{H}}\sim 10^{21}\ \mathrm{cm^{-2}}$, without considering any reflection or partial covering. In a variability study, \citet{Ri2011} also finds warm absorption and  estimated a lower limit for the location of the absorbing clouds, corresponding to the BLR.\\

\textbf{\it{Mrk 231}}\\

This is a well studied source and, among the many publications, \citet{Br2004} argues that the resulting flat photon index suggests a heavily absorbed spectrum consisting of a scattered power law component and a reflected component, and even a reflection dominated scenario. We do not consider any reflection components, but our results are consistent with a continuum well fitted by a power law with a flat slope and a column density of the order of $\sim 10^{22}\mathrm{cm^{-2}}$.\\

\textbf{\it{NGC 4593}}\\

Our results for this source are consistent with previous analysis, such as \citet{Br2007, E2013, U2016}. In particular, \citet{E2013}, worked with grating spectra and found 4 warm absorbers of different ionization states with at least one of high-ionization ($\log\xi \sim 2.5$) and determined its distance from the source of around a few pc. This result is also consistent with the work of \citet{U2016}, where they report two different warm absorbers, with the high-ionization one consistent with the column density and ionization parameter that we found; they determined this component to be at a distance of $\leq 3\ \mathrm{pc}$ from the central region.\\

\textbf{\it{ESO 323-G077}}\\

Our model does not accurately fit the continuum, possibly due to the omission of the reflection component and a soft excess that is modeled by more than one component. \citet{GM2013, JB2008} present a more detailed analysis. Both analysis find 2 warm absorbers as well as cold absorber that affects the power law in the hard band. \citet{GM2013} presents the idea that the variability observed in the warm absorber can be due to a clumpy torus or clouds in the BLR and proposes this to be a source of intermediate type between 1 and 2, being observed at an angle of $\sim 45^{\circ}$.\\

\textbf{\it{IRAS 15091-2107}}\\

\citet{JB2007} reports a cold and a warm absorber in this source. We find our best fitted model to be the one with a warm absorption since we do not consider a combination of both warm and cold absorption.\\

\textbf{\it{Fairall 51}}\\

This source shows significant improvement by adding a second warm absorber. \citet{S2015} presents an in-depth analysis on this source, where the best fit model includes a cold absorber $N_{\mathrm{H}}\sim4\times10^{22}\ \mathrm{cm^{-2}}$ affecting the hard band and the need of a reflection component. One important point on the absorbers is that they find an improvement on the fit by allowing the covering factor of the {\textit{zxipcf}} model to vary.\\

\subsubsection{Equatorial-polarized sources}

\textbf{\it{IZw 1}}\\

\citet{Ga2007, Cos2007} report on the spectral details and variability between two observations. They find evidence of absorption of two different ionization states. \citet{Si2018} worked on the same observation as we did and found a variable multi-phase ionized absorber by two gas components of similar column densities. The outflow velocity estimated for the low ionization component is of  $\sim 1900\ \mathrm{km/s}$  and for the high ionization is $\sim2500\ \mathrm{km/s}$. By assuming that the outflow velocity is greater or equal to the escape velocity, they estimate the absorbers distance to the central source at $\sim 0.07, 0.04\  \mathrm{pc}$ respectively, placing them in the scale of the accretion disk.\\

\textbf{\it{Akn 120}}\\

This source is refer to as a ``bare-nucleus" AGN, and thus, there are no reports of this source being affected by absorption. The spectra is characterized by a strong soft excess emission, as studied by \citet{Po2018} and \citet{Ma2014}.\\

\textbf{\it{NGC 3783}}\\

According to the AIC, we are unable to determine whether the warm or cold absorber yield a better fit for this source. There are many previous publications on this AGN including reports on obscuring events. Based on RGS data, \citet{Bl2002}, reported the detection of a two-phase warm absorber. This result is confirmed by \citet{Mao2019}.\\ 

\textbf{\it{Mrk 841}}\\

Our best fit for this source includes 2 warm absorbers of different ionization states, a result that is consistent with the in depth analysis made by \citet{AL2010}. In this work, they estimate the density of the absorbing gas below $10^{3}\mathrm{cm^{-3}}$, and the distance to the central source of a few tens of pc, placing the absorber in the scale of the BLR.\\

\textbf{\it{Mrk 876}}\\

We report this source as unabsorbed. This result is consistent with work reported in \citet{Bo2022}, using {\it NuSTAR} data, and \citet{Bo2015} with {\it XMM-Newton} and {\it Swift}. This unabsorbed scenario corresponds to a classical Seyfert 1. \\

\textbf{\it{KUV 18217+6419}}\\

The best fit from our analysis indicates the presence of a warm absorber. This source shows interesting features in the hard band, \citet{JB20072}, but we found no studies on absorption in the soft band.\\ 

\textbf{\it{Mrk 509}}\\

This source has been largely studied, in particular in a campaign lead by \citet{K2011}, where we can find reports on the many components of the spectrum. In particular, \citet{De2011} finds multiple absorption systems, with three different velocity components. This is not consistent with our results, where this source appears unabsorbed, according to the AIC. This is an example of how an over-simplified model like ours does not recover the spectral complexity of the source.\\

\textbf{\it{Mrk 304}}\\

In our work, the AIC favors the cold absorption scenario. Our result differs from previous studies of the same observation. We find a cold absorber of $\sim 10^{21}\ \mathrm{cm^{-2}}$ and our baseline model does not reproduce the convex shape of the continuum. \citet{Pi2004, JB2004} find that the convex shape of the spectra corresponds to a heavily obscured source with column density up to $\sim 10^{23}\ \mathrm{cm^{-2}}$, and their final fit includes a multi-phase ionized absorber.\\

\section{Conclusions} \label{sect:Conc}

We presented the results of a systematic analysis of the X-ray spectra of 19 sources: 11 with \textit{PL-pol} and 8 with \textit{EQ-pol}. Our analysis consisted of fitting the main components of a typical Seyfert spectrum and testing the response when absorption was added to the model. We particularly tested whether cold or warm absorption was a preferred solution, as determined by the AIC statistical criterion.

\begin{itemize}

    \item Concerning the common components of the spectral modeling, we find that the continuum is well-fitted by a power law, the $\mathrm{Fe\ K\alpha}$ is present in 13 out of the 19 observations, and the soft excess is ubiquitous. These components prove to be a good first approximation to model the typical type 1 Sy spectra. However, comparing our results to previously published work on these sources, we find that  producing a more robust best fit model for the absorption components requires a more detailed analysis, to be complemented by  high-resolution data. This suggests that our study can be further improved by the selection of a more complex model.
    
    \item From the absorption test, we find that $100\%$ of the \textit{PL-pol} and $75\%$ of the \textit{EQ-pol} sources are affected by absorption. This difference, which is corroborated by the statistical KS test, seems to indicate an intrinsic diversity of the scattering medium in the two groups of sources, lending support to the unification model proposed by Smith. 

    \item ``While we observe a distinction between sub-samples, another interesting result emerges when examining the entire sample of polarized Sy 1. The incidence of absorption in 19 type 1 Sy is of $\sim 90\%$, with $\sim 75\%$ confirmed to be warm absorption. In contrast with results where warm absorption is found in $65\%$ of type 1 Sy, \citep{L2014I}, this suggests a relationship between the presence of absorbing material in the line-of-sight and the region responsible for the scattering that yields the measured optical polarization.

    \item At first approximation, our work provides a promising test for the use of X-ray absorption as a tool for explaining the properties of the observed polarization and their interpretation in the context of the AGN unification model. It is desirable that further work can include more recent polarization measurements and a larger sample of X-ray sources with known polarization. The test of X-ray absorption can also be improved by considering more complex modeling of the absorbers, aiming at constraining the location of the absorbing gas with respect to the AGN torus. It is important noting that, while the model presented by Smith can be regarded as a first approximation to a unification scheme, the availability and quality of X-ray data and therefore our understanding of the X-ray emission of AGN has significantly improved since Smith's unification scheme was first proposed (2002-2005). This poses the interesting option of taking this systematic analysis further by considering a multi-layered absorber, a less homogeneous torus and a more detailed characterization of the absorbing gas.

\end{itemize}

\bibliography{ref.bib}

\section{Acknowledgments}

 This research has been founded by the project PID2019-107408GB-C41 and PID2022-136598NB-C33 by the Spanish Ministry of Science and Innovation/State Agency of Research MCIN/AEI/10.13039/501100011033 and by ``ERDF A way of making Europe". MGY acknowledges ``Programa de Apoyo a los Estudios de Posgrado (PAEP), UNAM".
A.L.L. and MGY acknowledge support from DGAPA-PAPIIT grant IA101623.
A.L.L. acknowledges the staff of the European Space Astronomy Centre (ESAC, Madrid) for hosting her visit during which this work was substantially discussed and advanced. Financial support is acknowledged from ESA through the Science Faculty - Funding reference ESA-SCI-SC-LE-123,  and from project PID2019-107408GB-C41 by the Spanish Ministry of Science and Innovation/State Agency of Research MCIN/AEI/ 10.13039/501100011033.

\appendix 
\label{App:spec}

\section{X-ray Spectral Analysis}

\centering
\begin{figure}
\begin{adjustbox}{width=\textwidth}
\begin{tabular}{ccc}
  \includegraphics[scale=0.65]{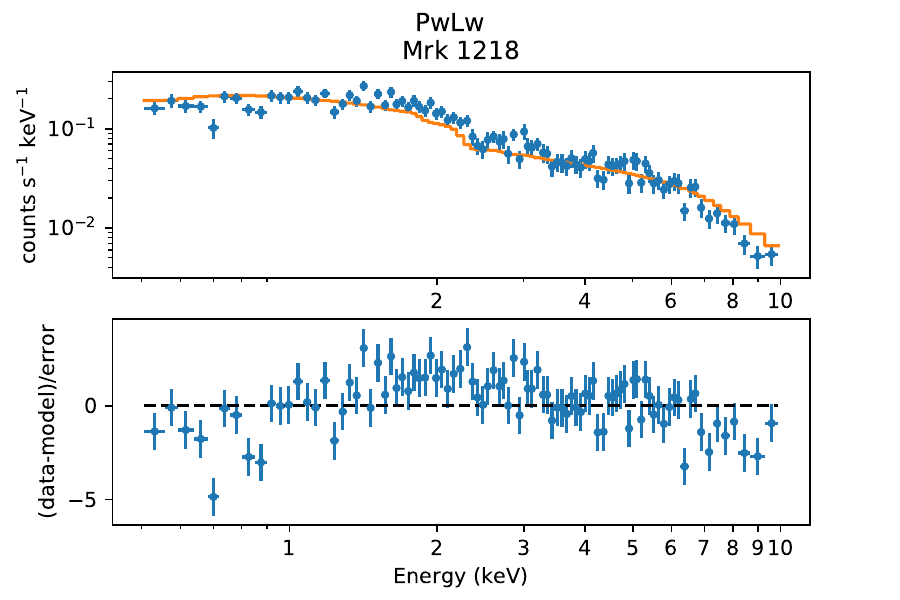} & \includegraphics[scale=0.65]{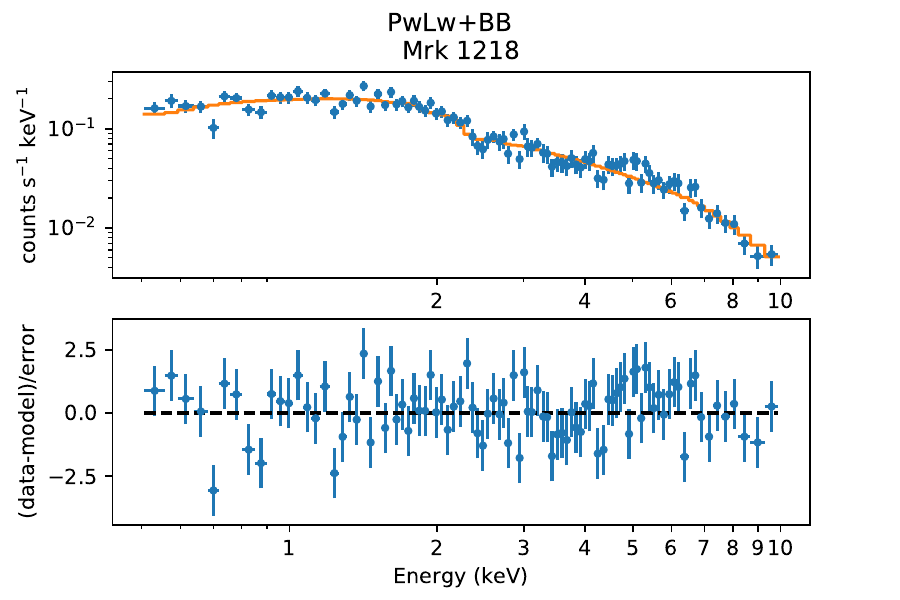} & \includegraphics[scale=0.65]{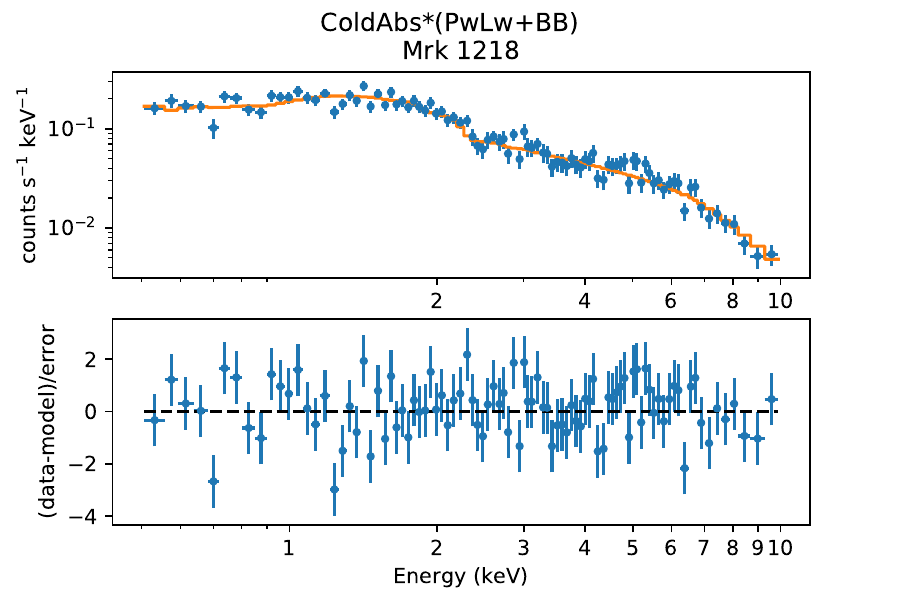}\\
  \includegraphics[scale=0.65]{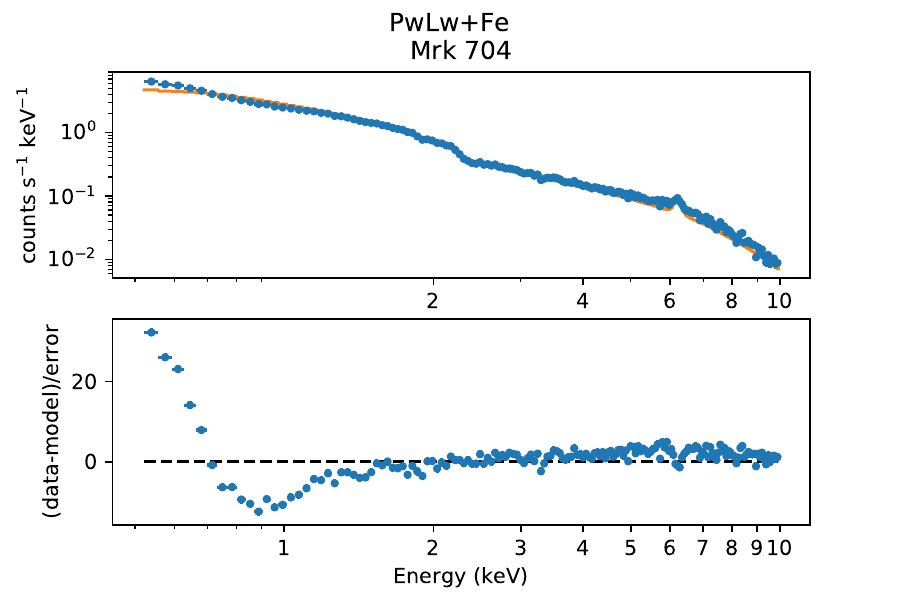} & \includegraphics[scale=0.65]{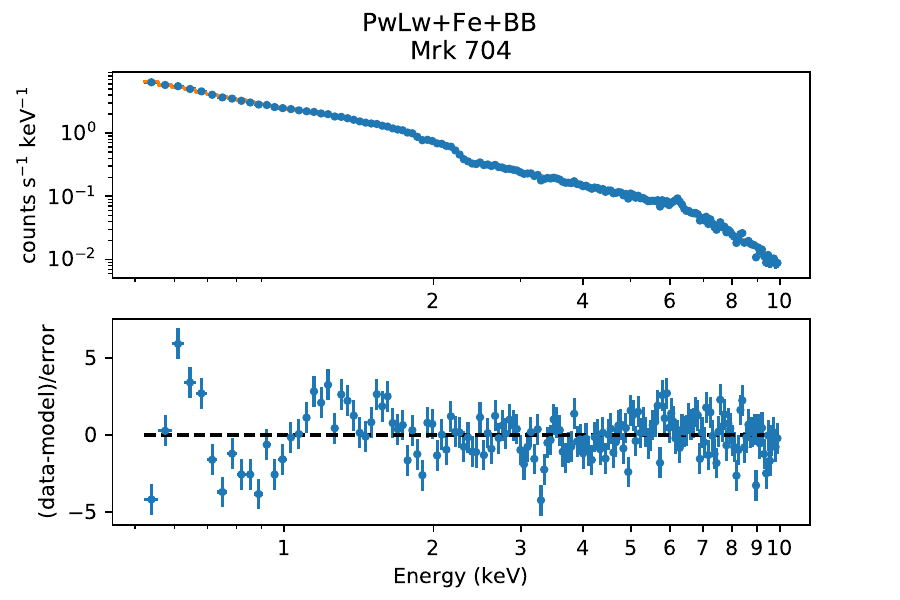} & \includegraphics[scale=0.65]{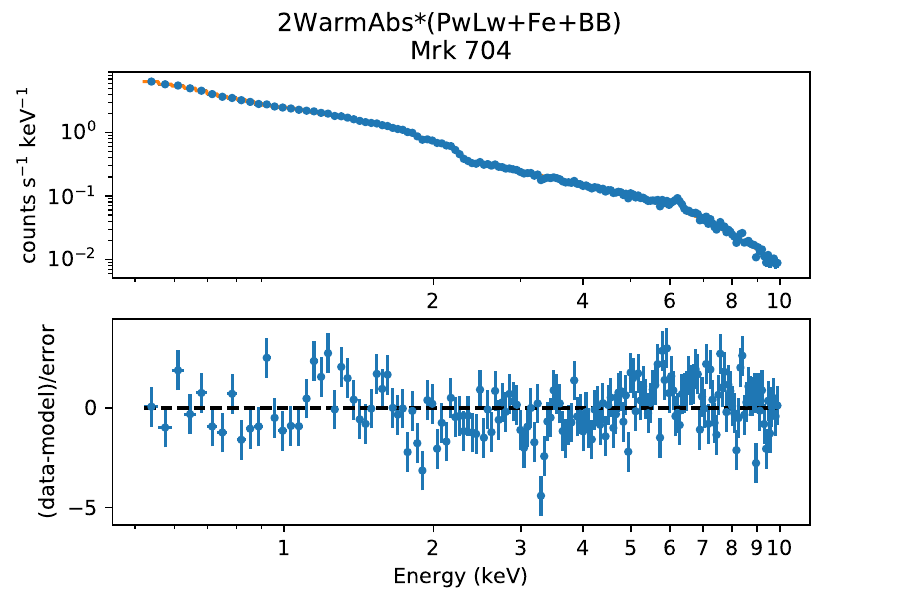}\\
  \includegraphics[scale=0.65]{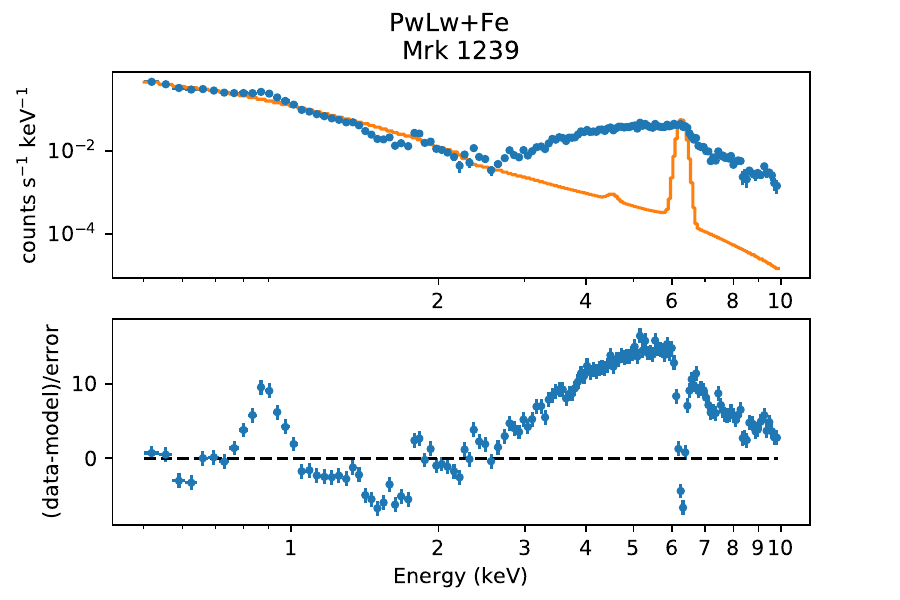} & \includegraphics[scale=0.65]{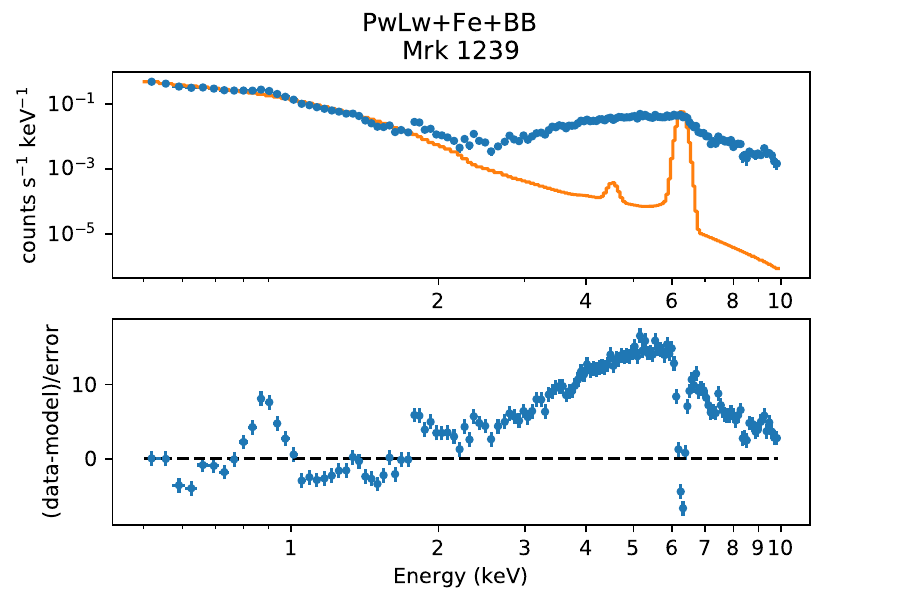} & \includegraphics[scale=0.65]{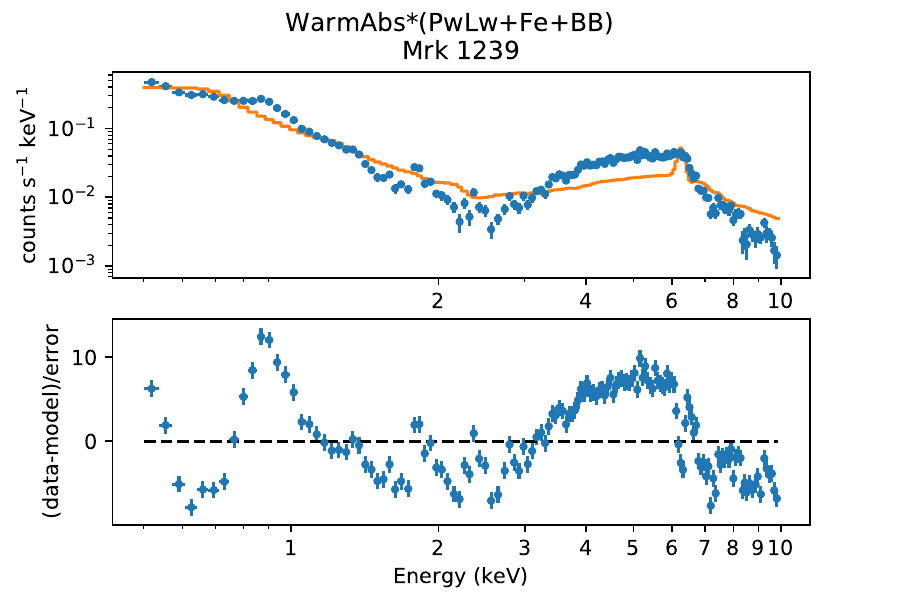} \\
  \includegraphics[scale=0.65]{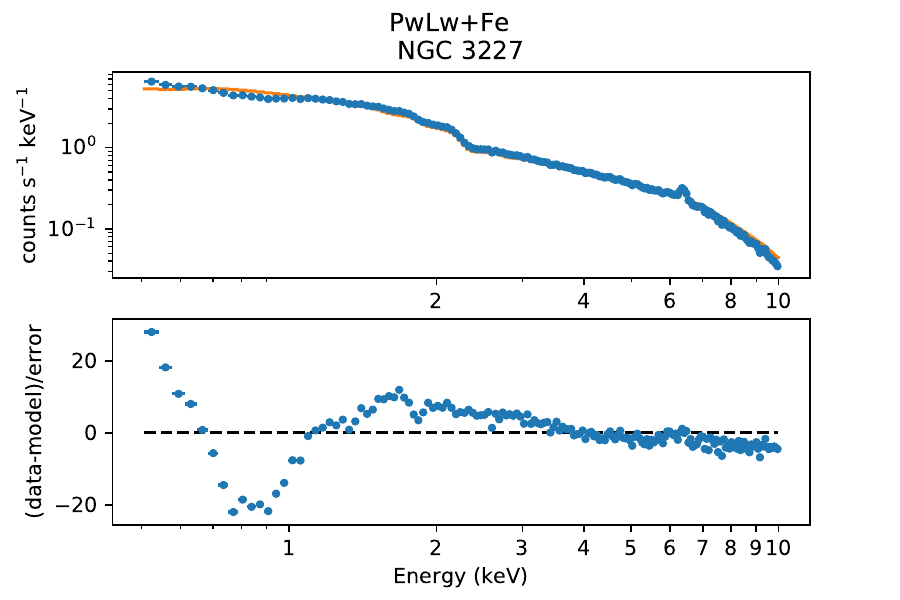} & \includegraphics[scale=0.65]{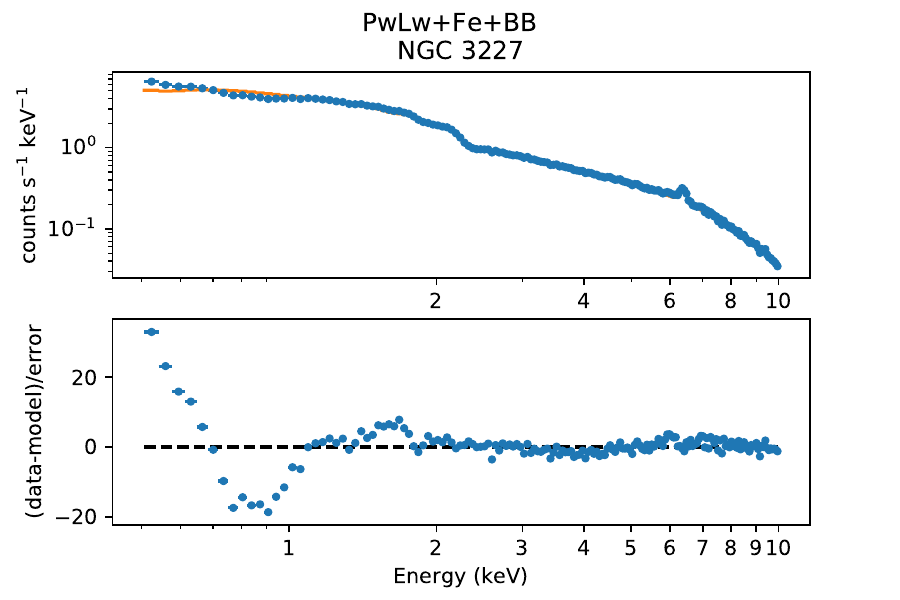} & \includegraphics[scale=0.65]{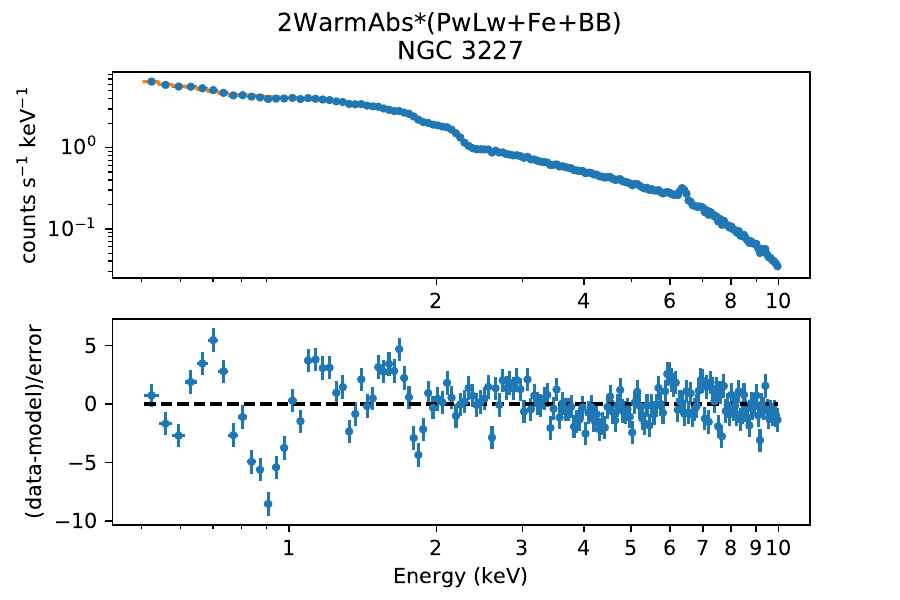} \\
  \includegraphics[scale=0.65]{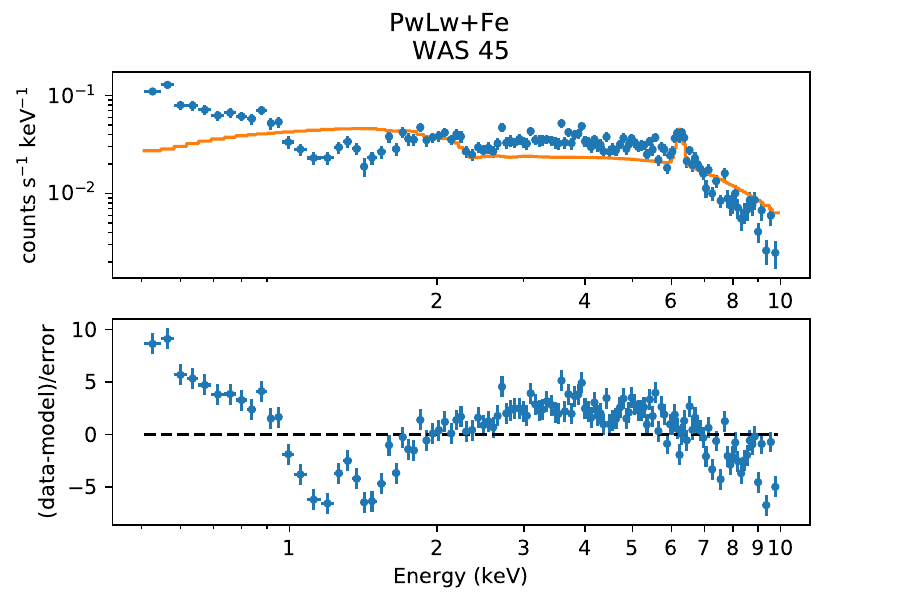} & \includegraphics[scale=0.65]{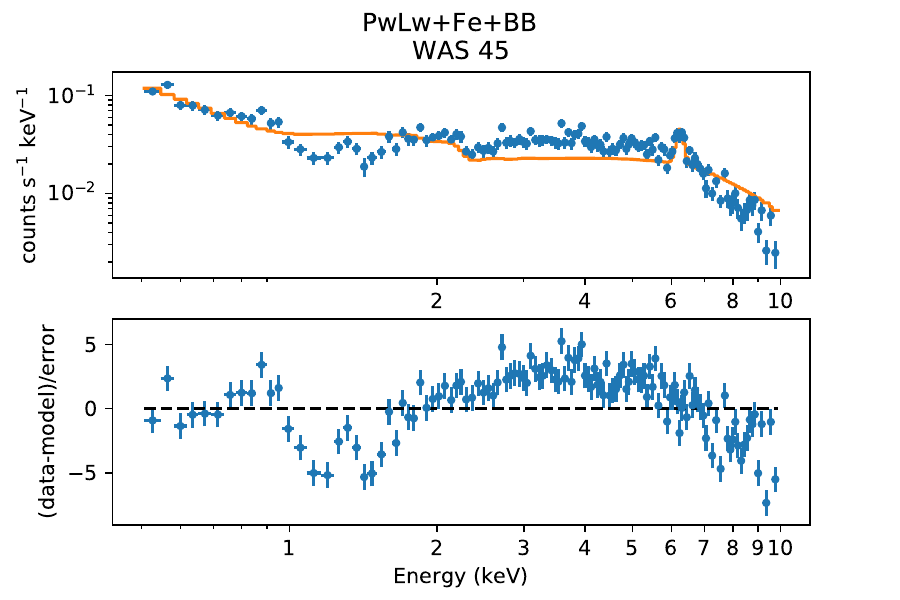} & \includegraphics[scale=0.65]{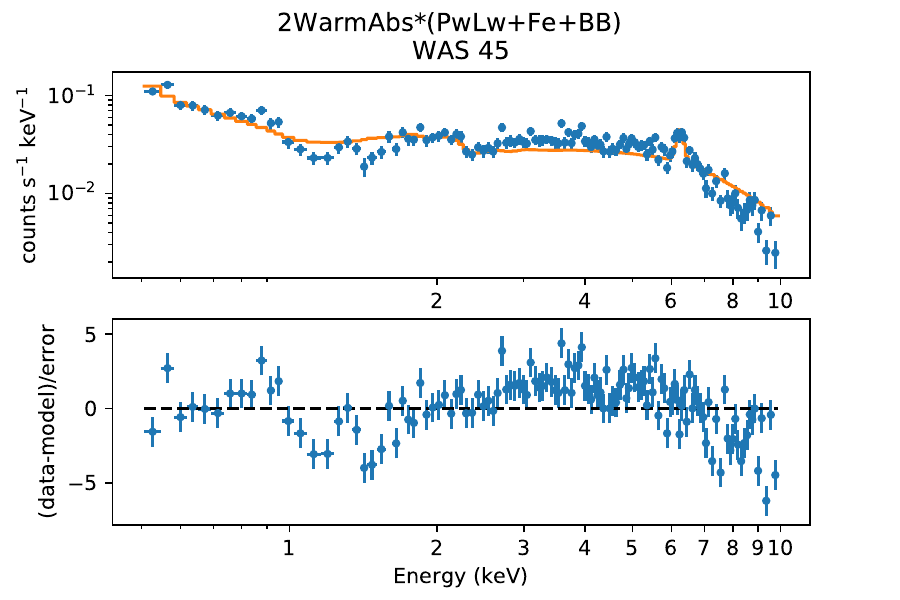} \\
\end{tabular}
\end{adjustbox}
\end{figure}
  
\begin{center}
\begin{figure}
\begin{adjustbox}{width=\textwidth}
\begin{tabular}{ccc}
  \includegraphics[scale=0.65]{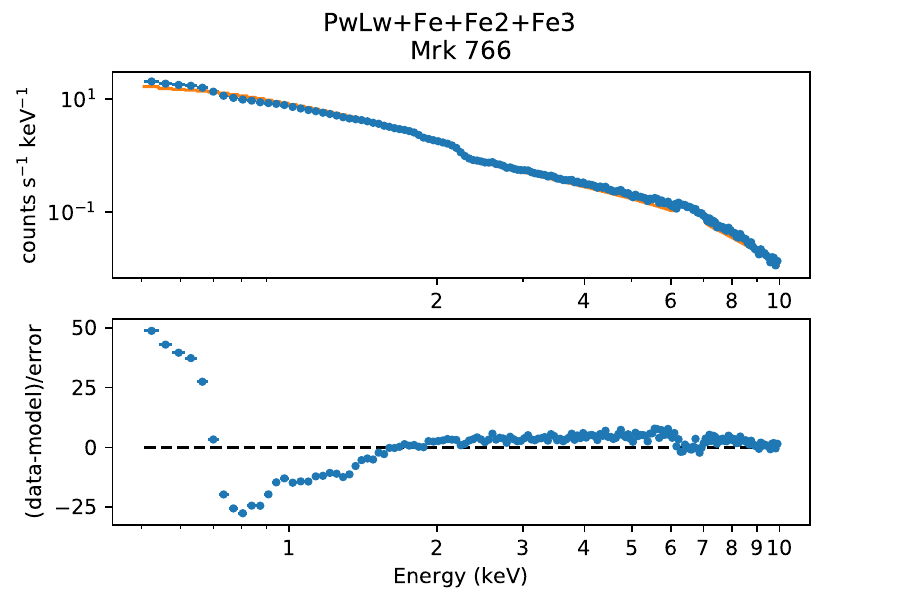} & \includegraphics[scale=0.65]{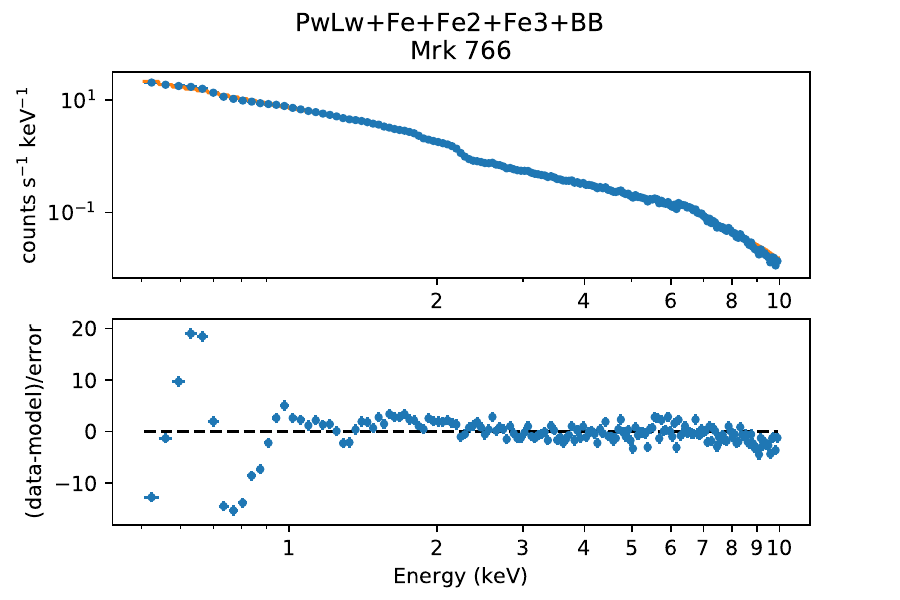} & \includegraphics[scale=0.65]{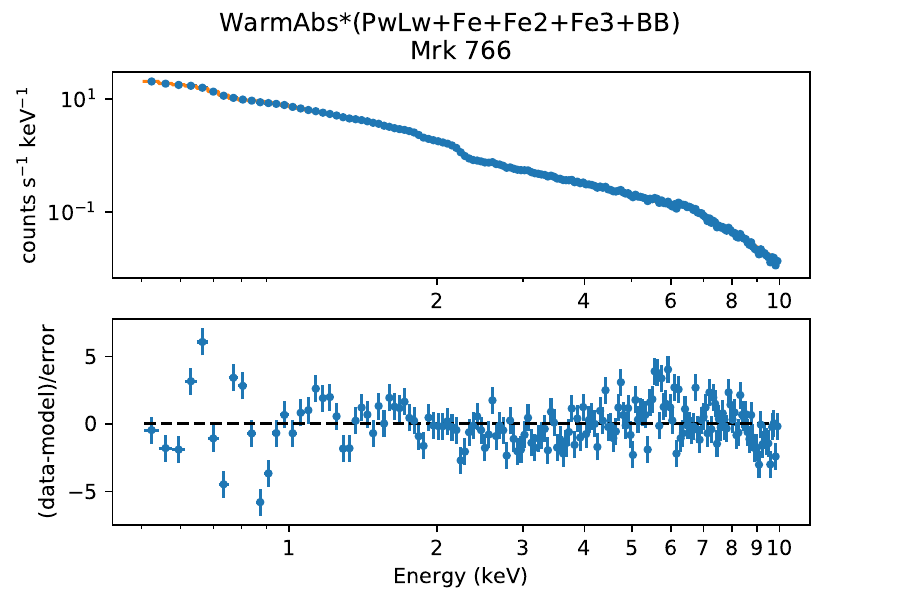} \\
  \includegraphics[scale=0.65]{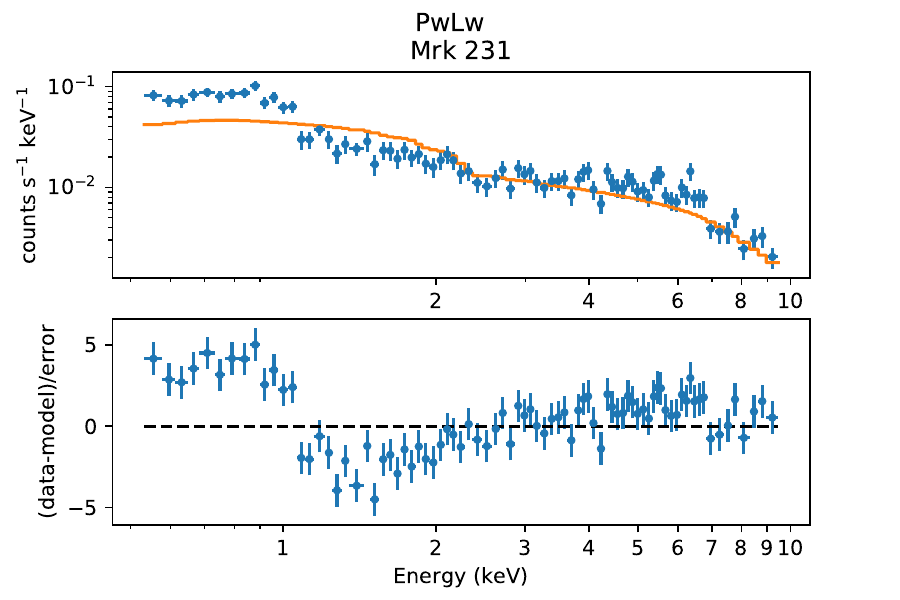} & \includegraphics[scale=0.65]{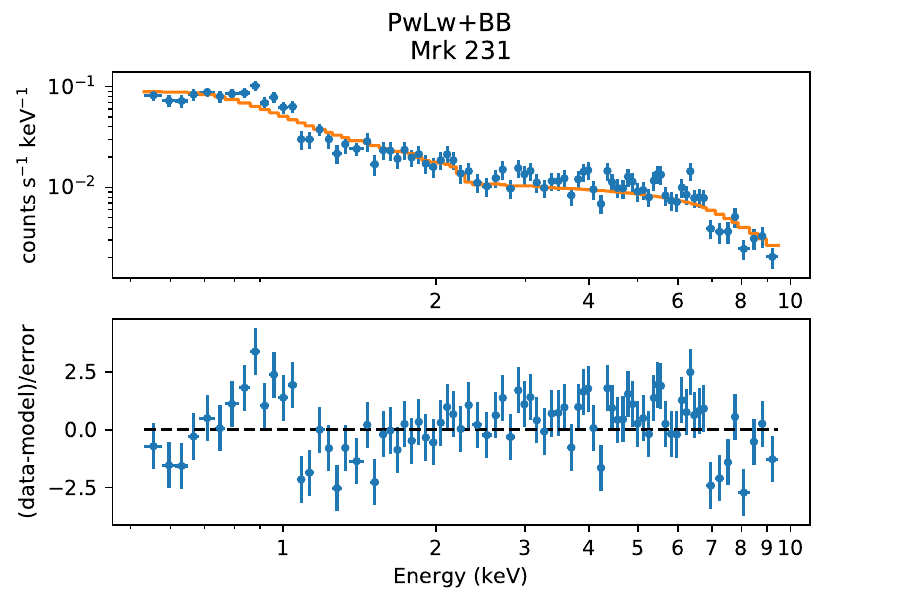} & \includegraphics[scale=0.65]{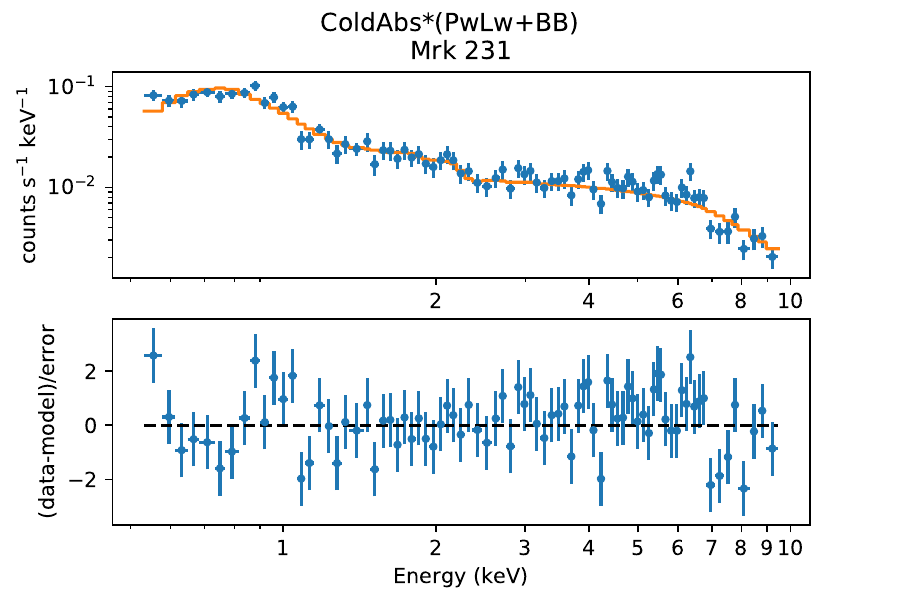} \\
  \includegraphics[scale=0.65]{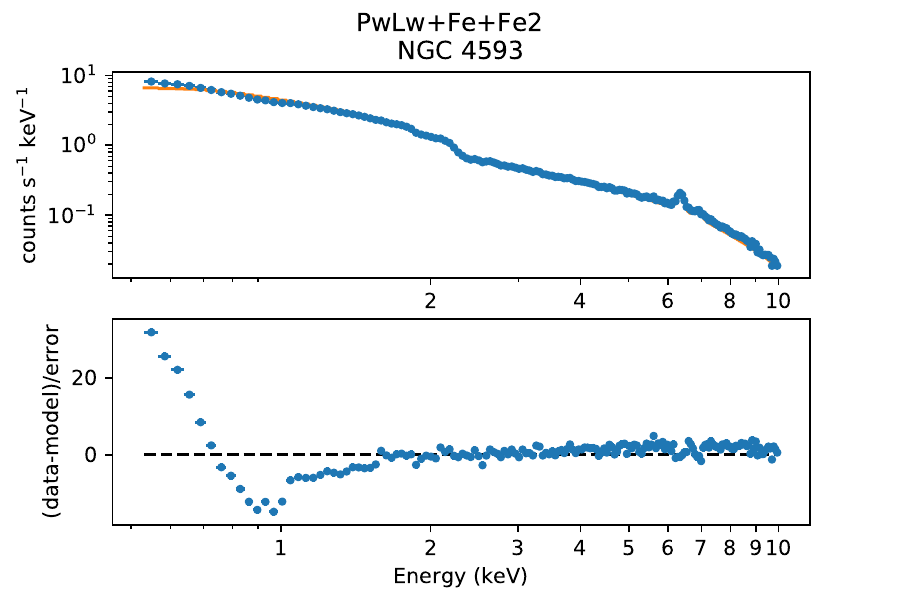} & \includegraphics[scale=0.65]{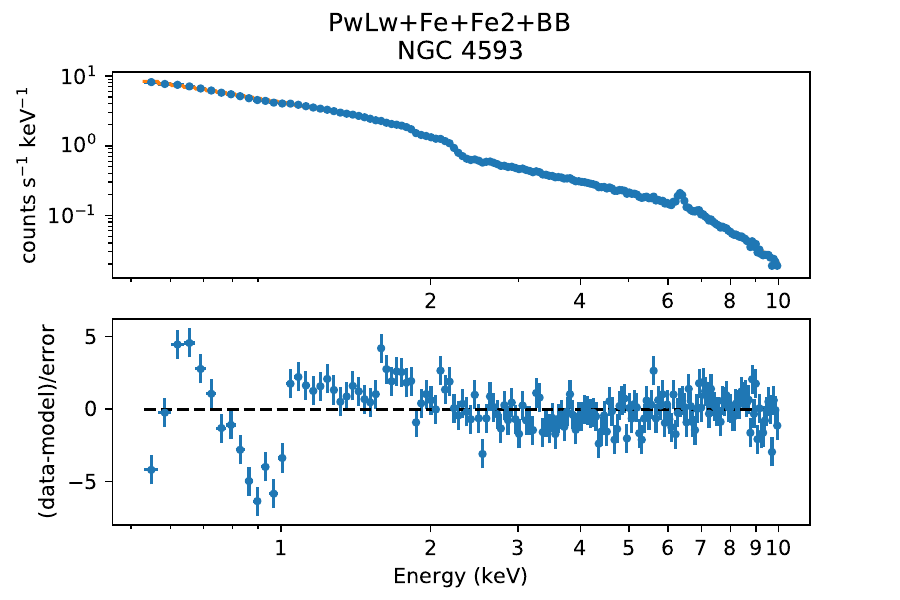} & \includegraphics[scale=0.65]{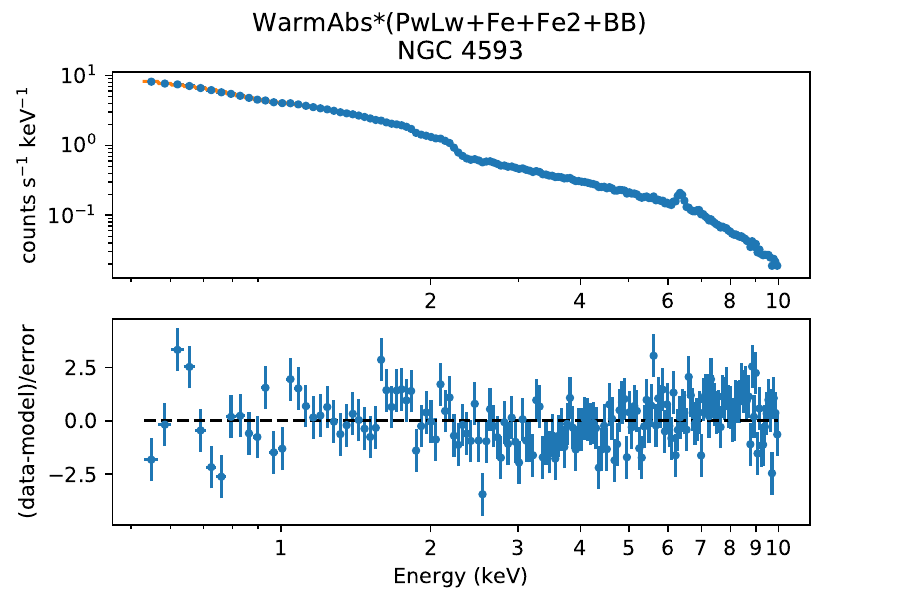} \\
  \includegraphics[scale=0.65]{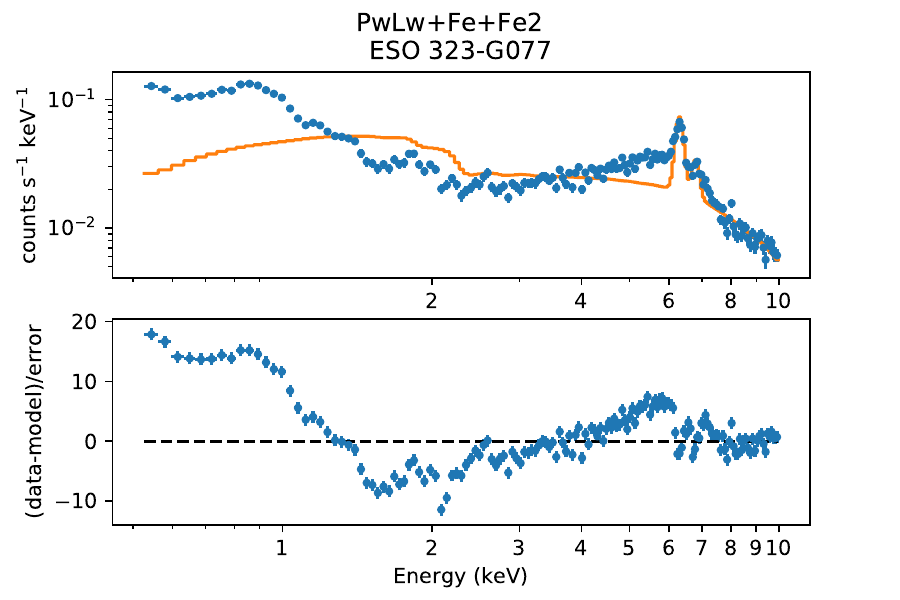} & \includegraphics[scale=0.65]{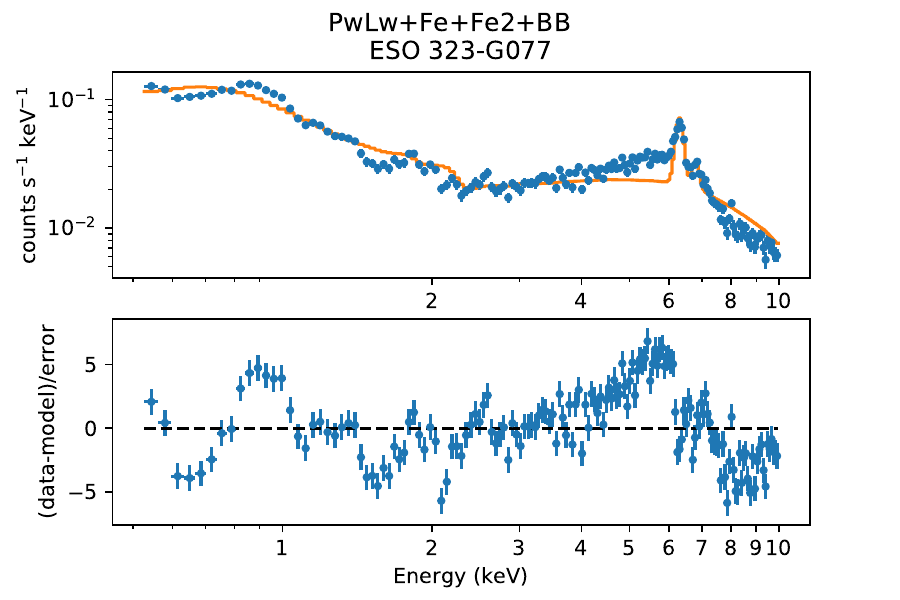} & \includegraphics[scale=0.65]{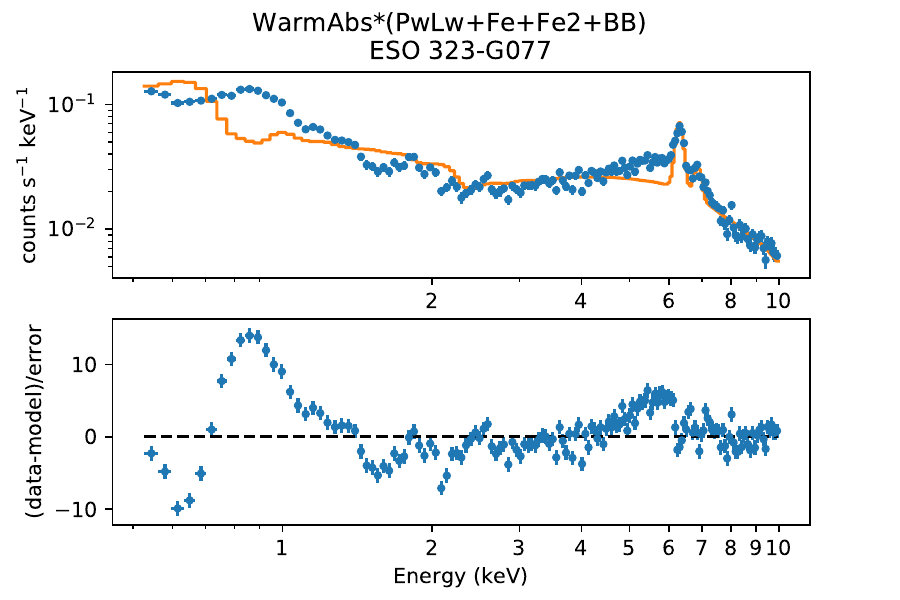} \\
  \includegraphics[scale=0.65]{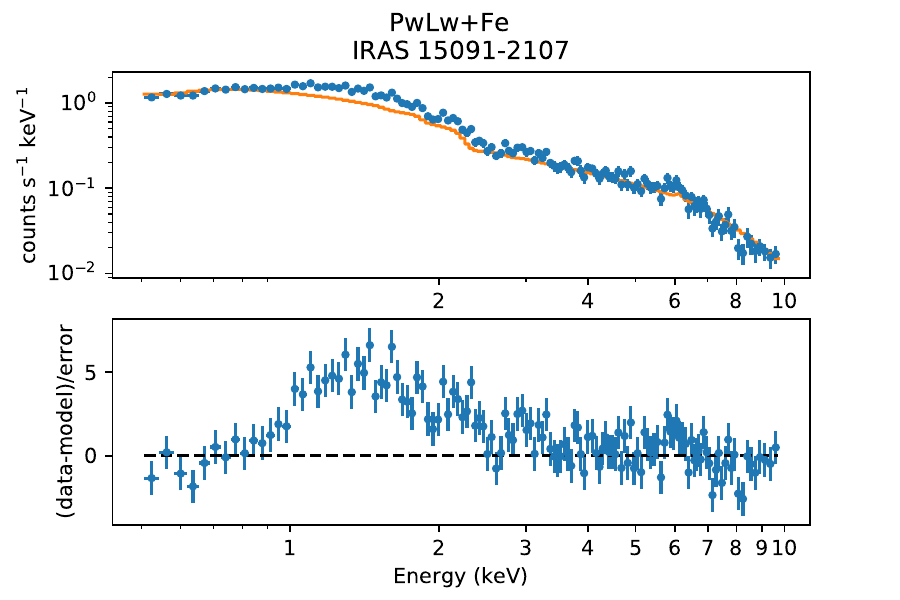} & \includegraphics[scale=0.65]{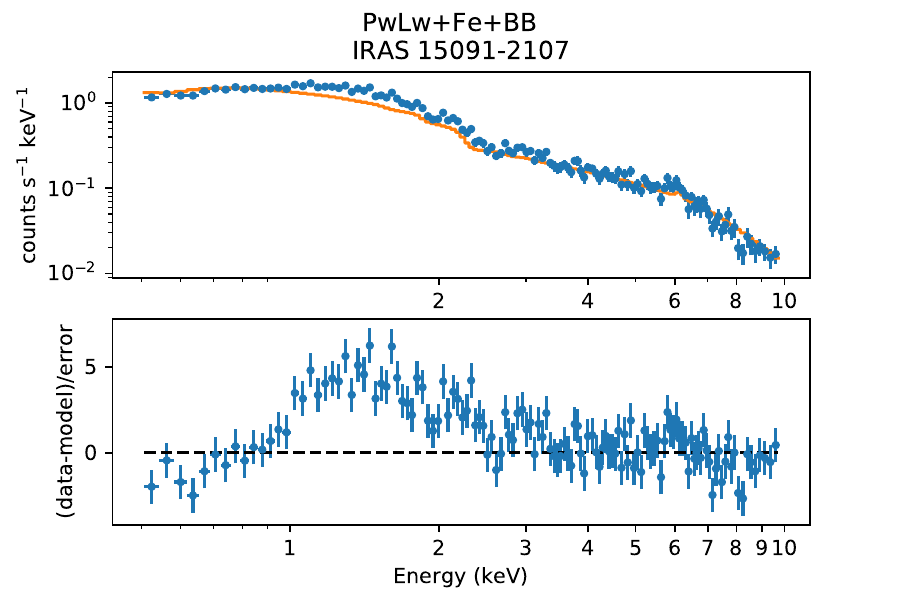} & \includegraphics[scale=0.65]{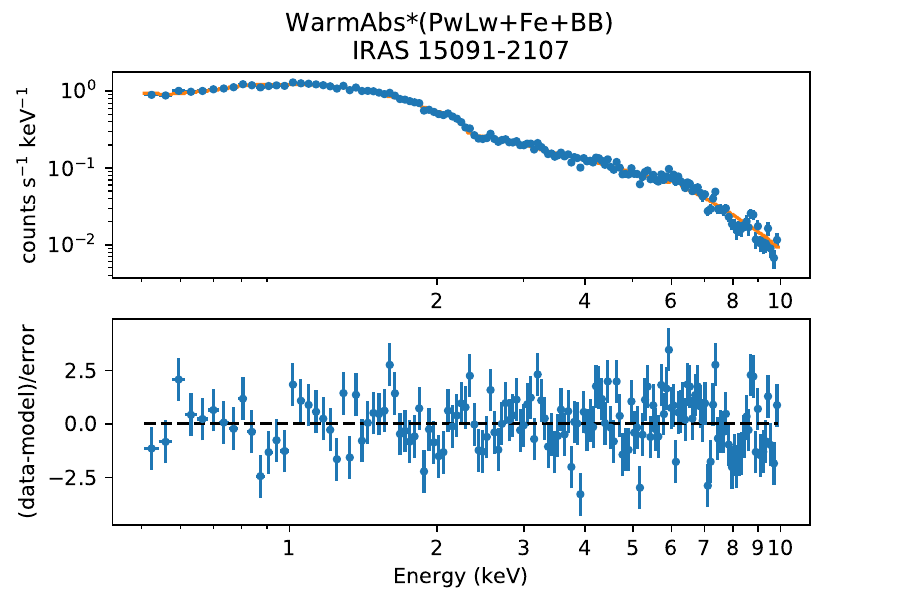} \\
  \includegraphics[scale=0.65]{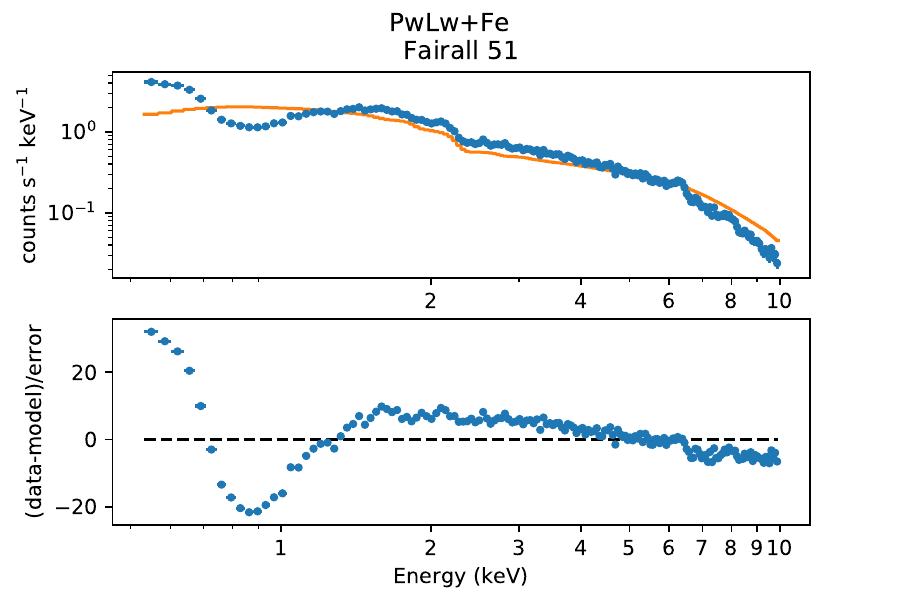} & \includegraphics[scale=0.65]{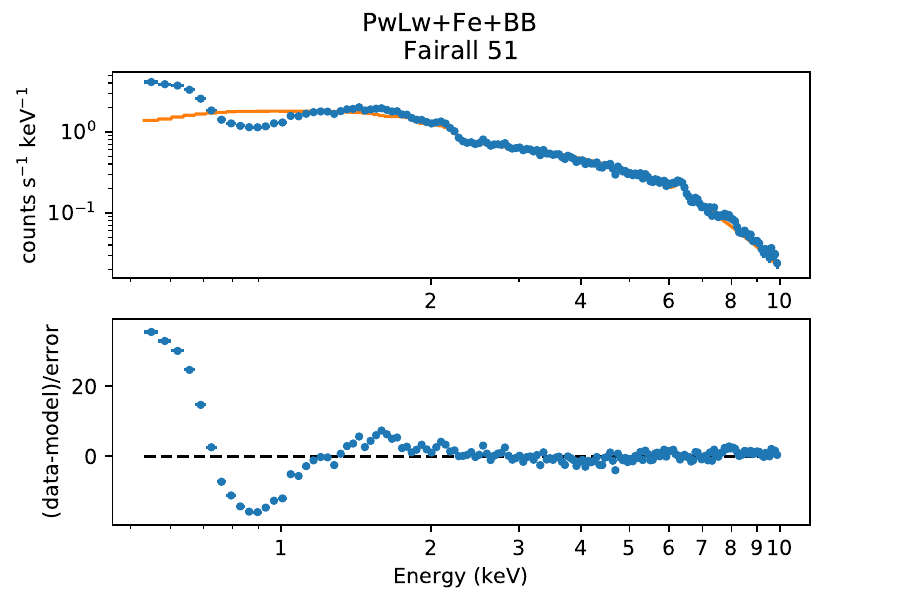} & \includegraphics[scale=0.65]{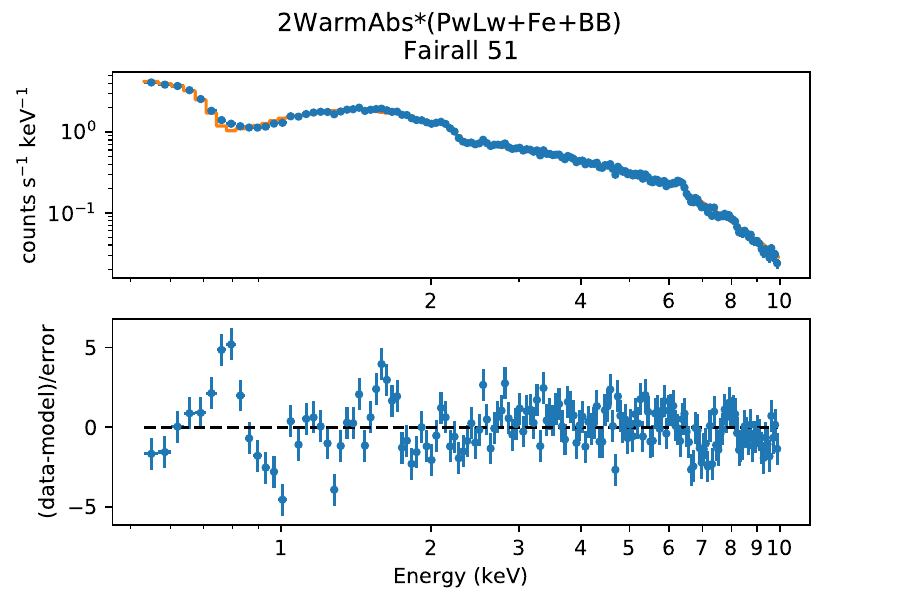} \\
\end{tabular}
\end{adjustbox}
\caption{\textbf{Polar polarized sources}. All the spectra are fitted in a range of 0.5-10 keV. First column corresponds to the baseline model: power law + significant Fe emission lines. Second column are fits with the addition of soft excess as black body. Third column is the resulting model of the absorption test, either cold or warm.}
\end{figure}
\end{center}

\begin{figure}
\begin{adjustbox}{width=\textwidth}
\begin{tabular}{ccc}
    \includegraphics[scale=0.65]{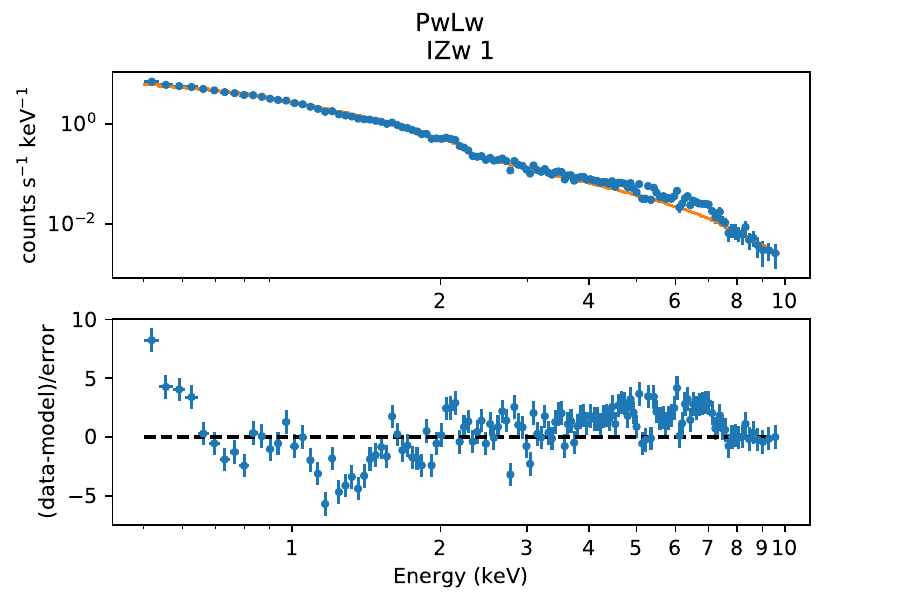} & \includegraphics[scale=0.65]{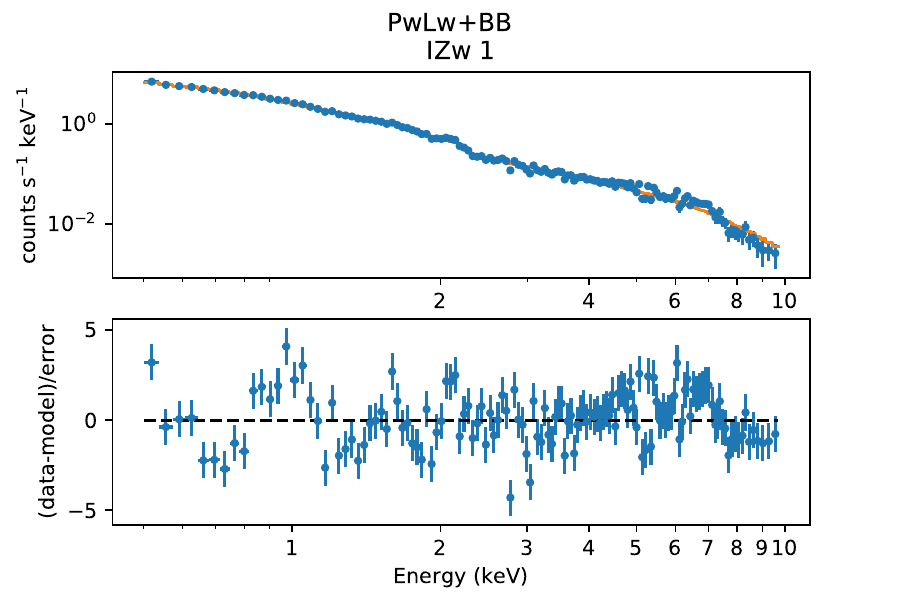} & \includegraphics[scale=0.65]{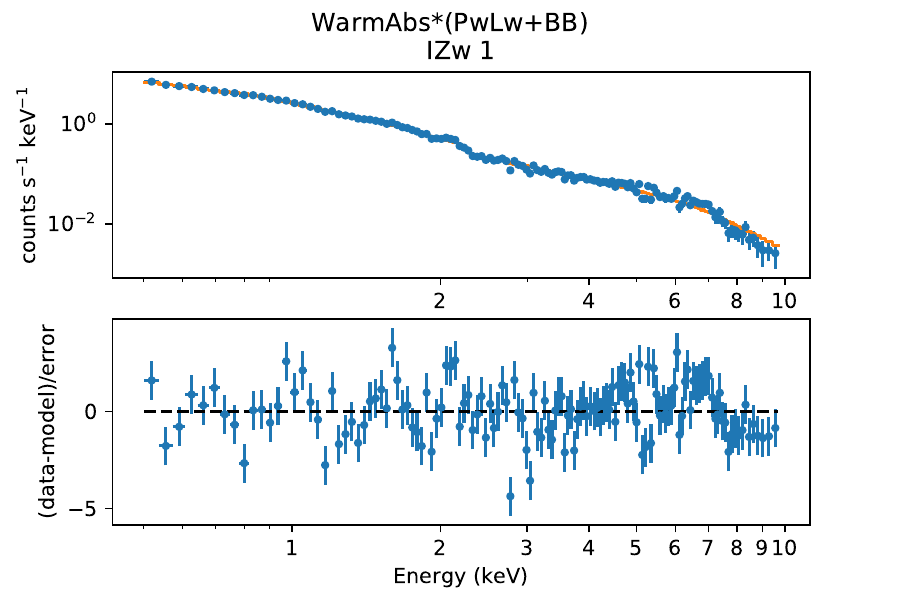} \\
    \includegraphics[scale=0.65]{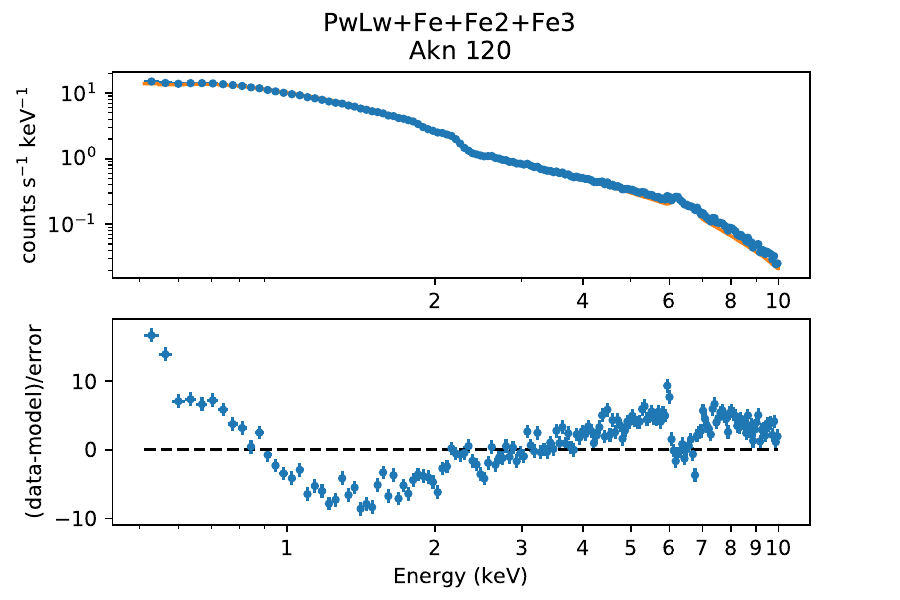} & \includegraphics[scale=0.65]{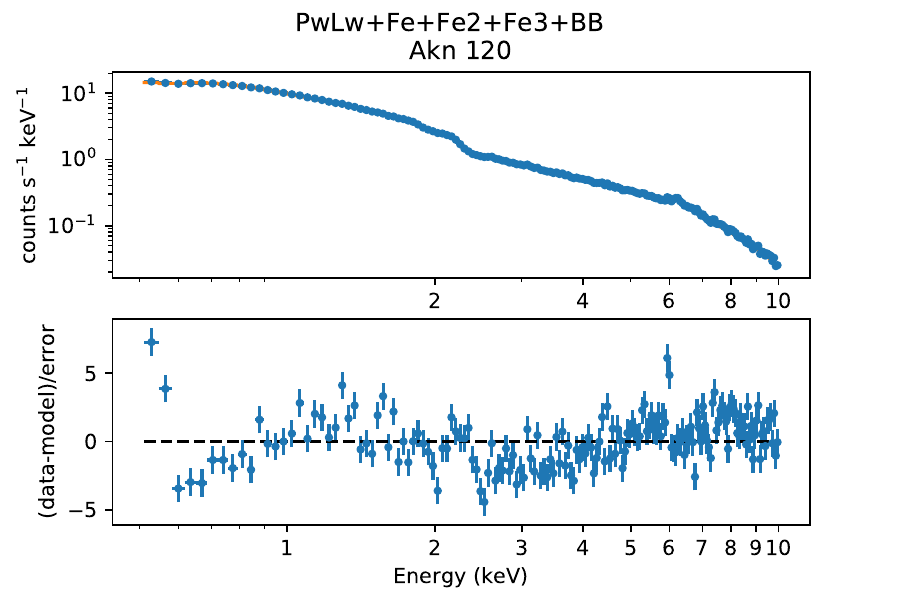} & \includegraphics[scale=0.65]{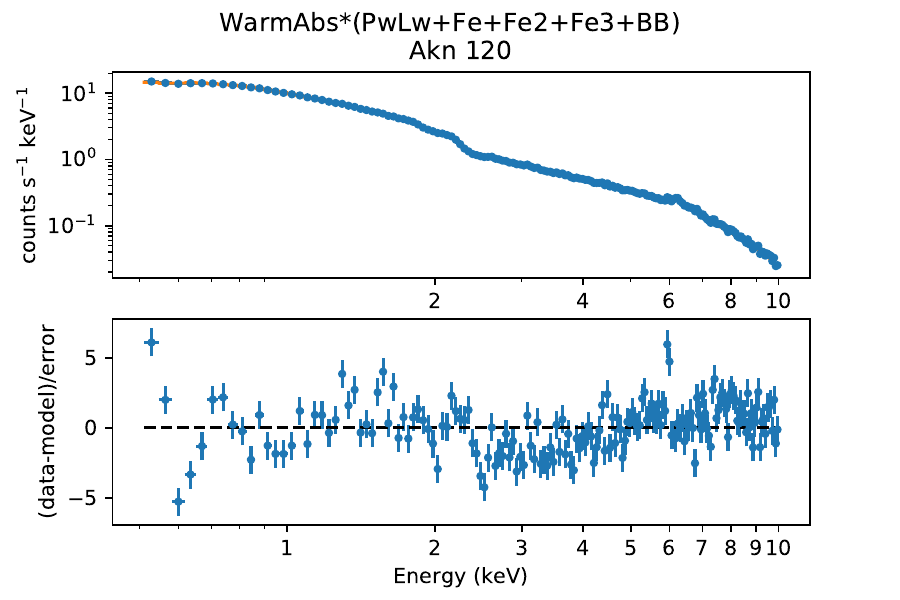} \\
    \includegraphics[scale=0.65]{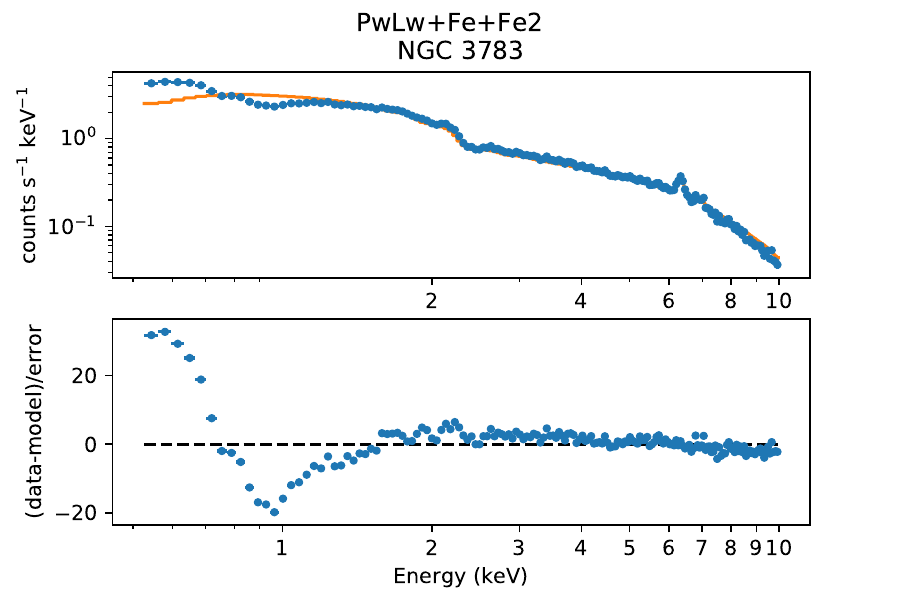} & \includegraphics[scale=0.65]{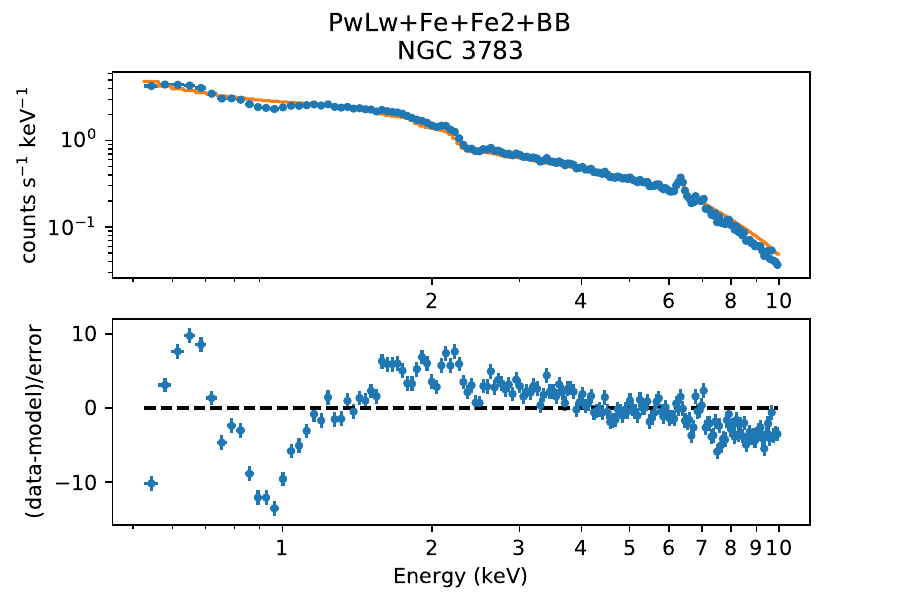} & \includegraphics[scale=0.65]{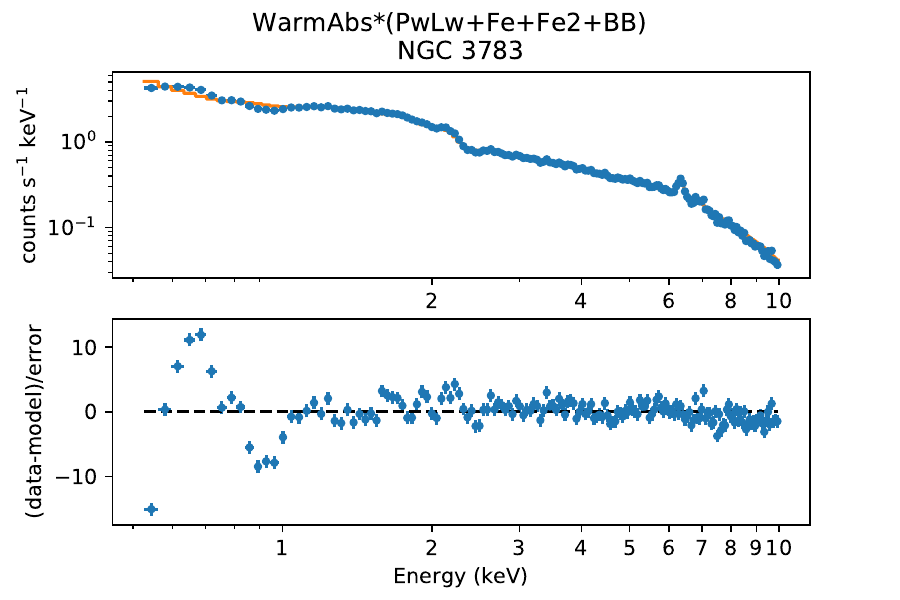} \\
    \includegraphics[scale=0.65]{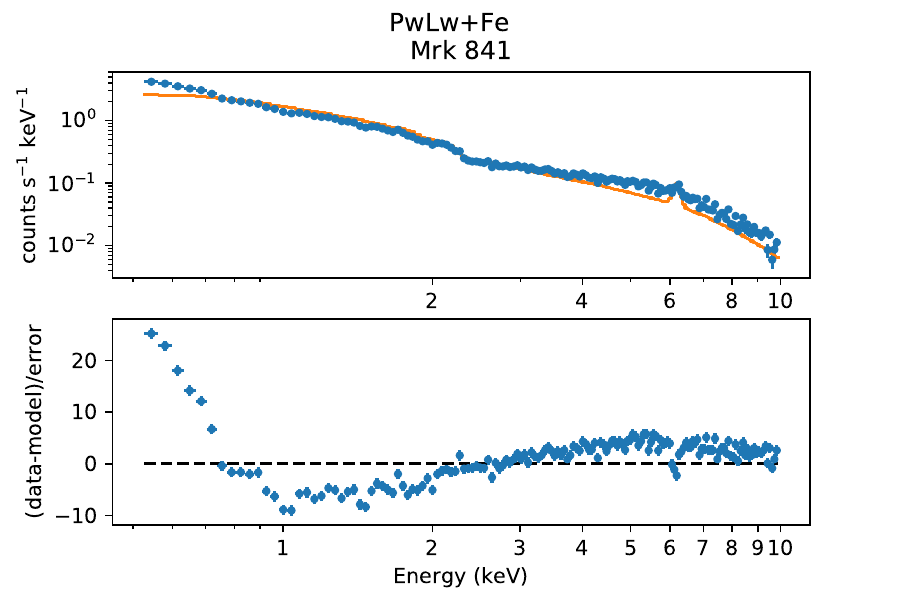} & \includegraphics[scale=0.65]{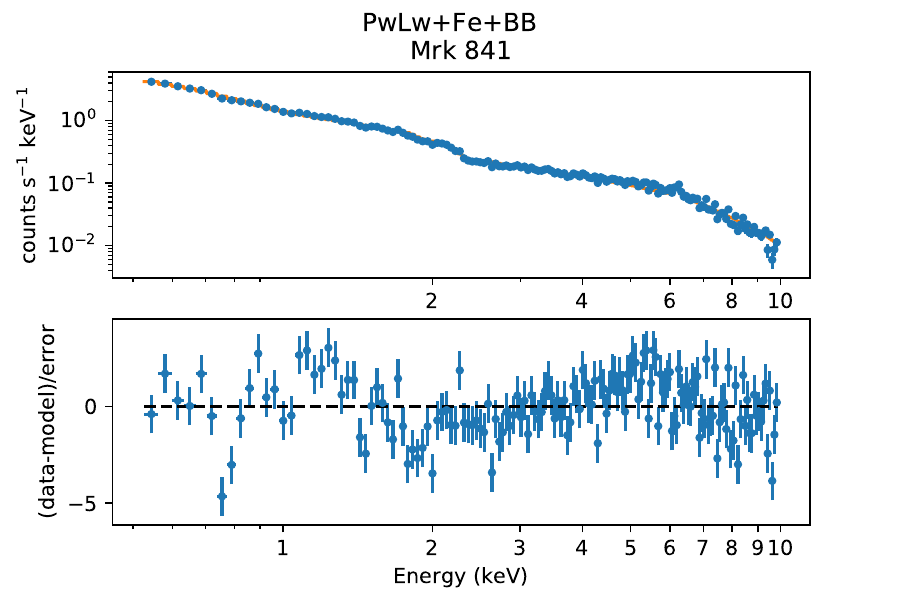} & \includegraphics[scale=0.65]{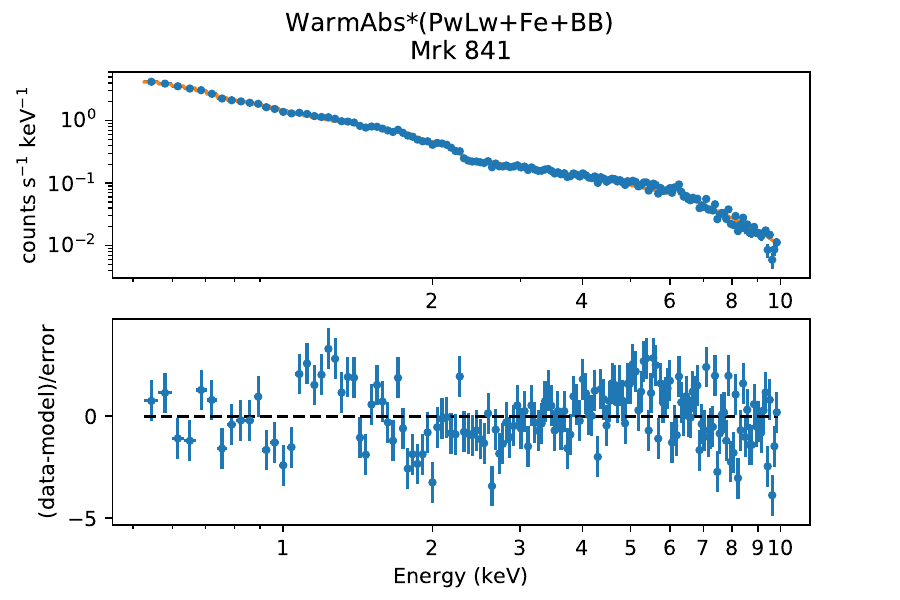} \\
    \includegraphics[scale=0.65]{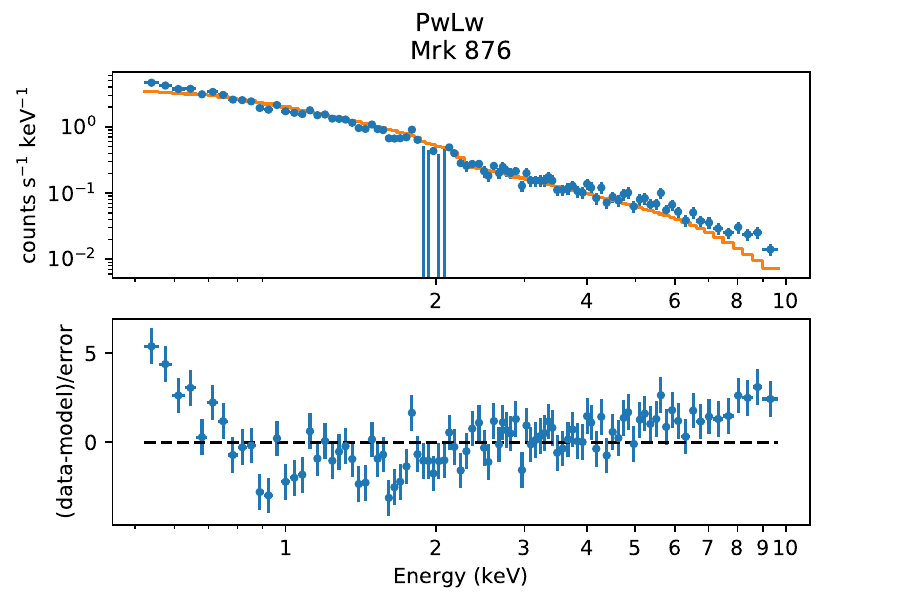} & \includegraphics[scale=0.65]{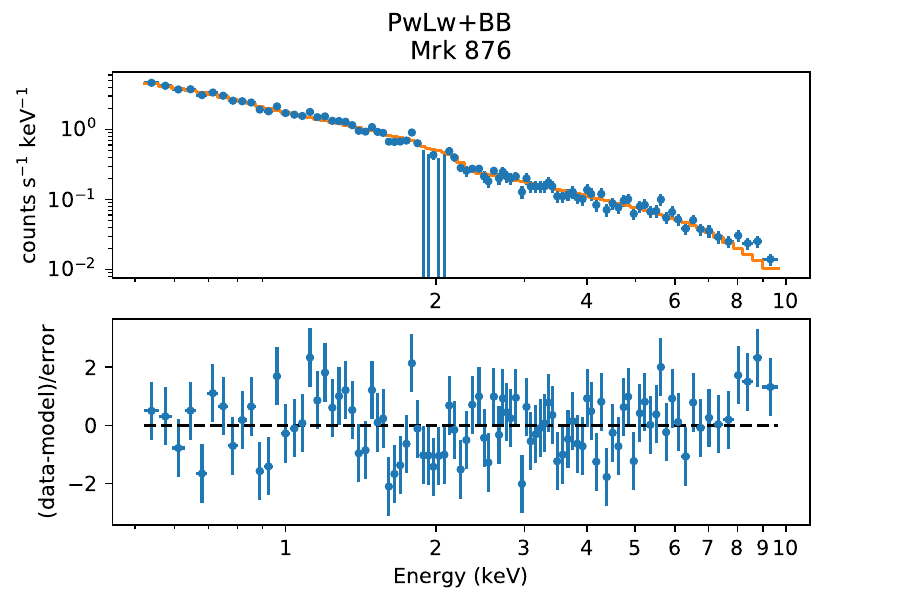} & \includegraphics[scale=0.65]{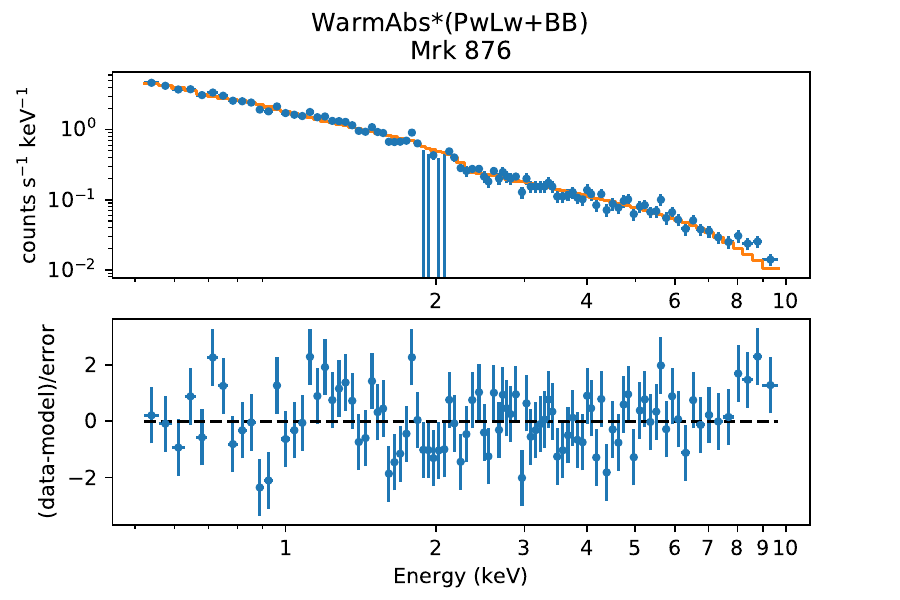} \\
    \includegraphics[scale=0.65]{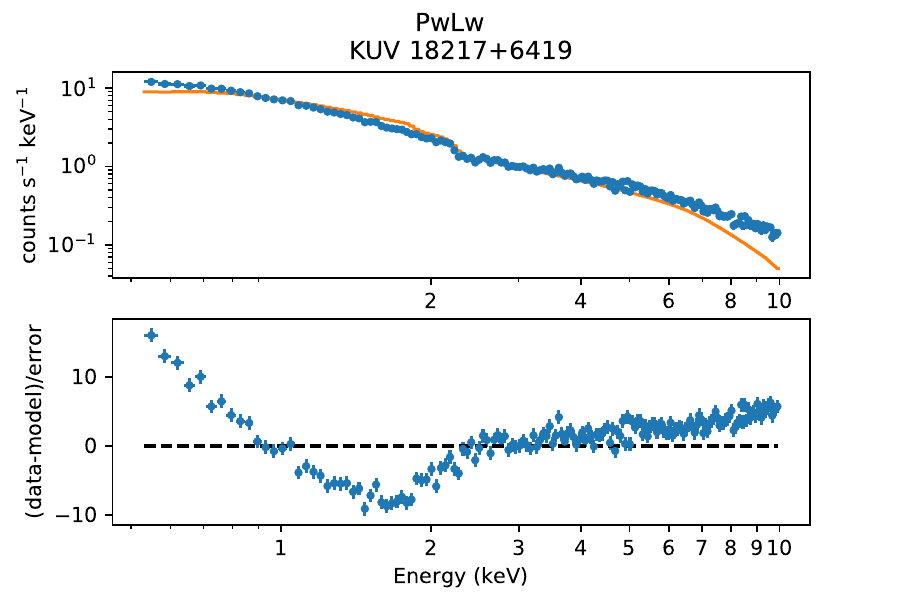} & \includegraphics[scale=0.65]{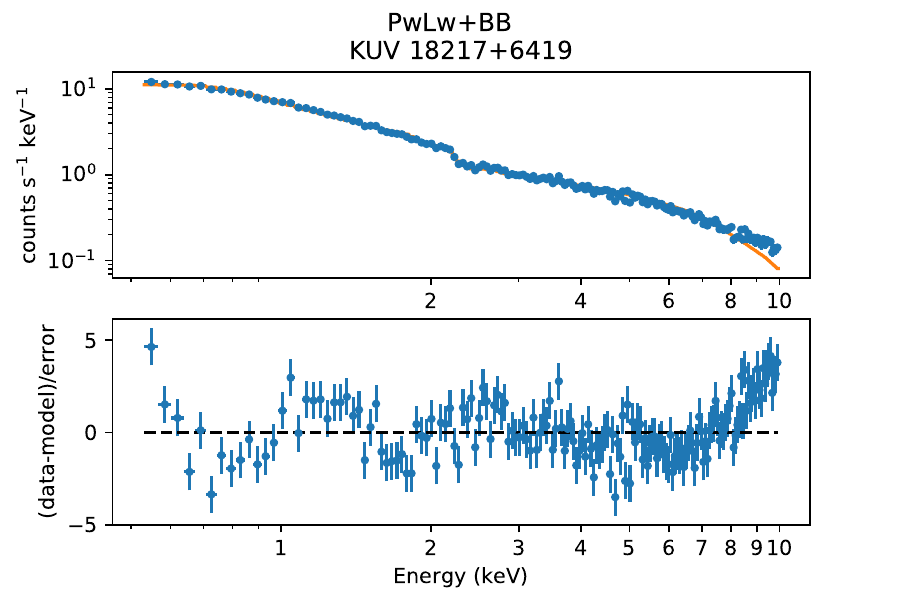} & \includegraphics[scale=0.65]{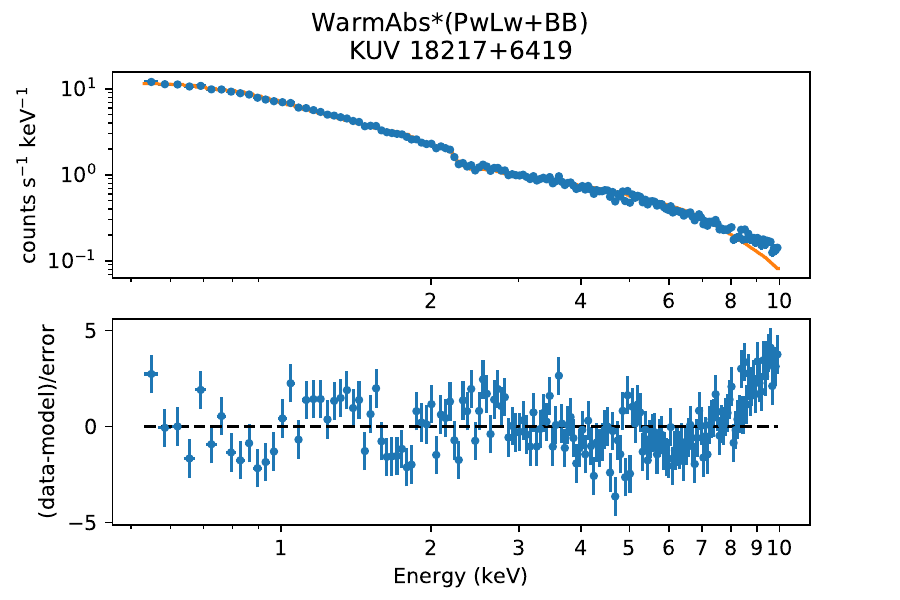} \\
\end{tabular}
\end{adjustbox}
\end{figure}

\begin{figure}
\begin{adjustbox}{width=\textwidth}
\begin{tabular}{ccc}    
    \includegraphics[scale=0.65]{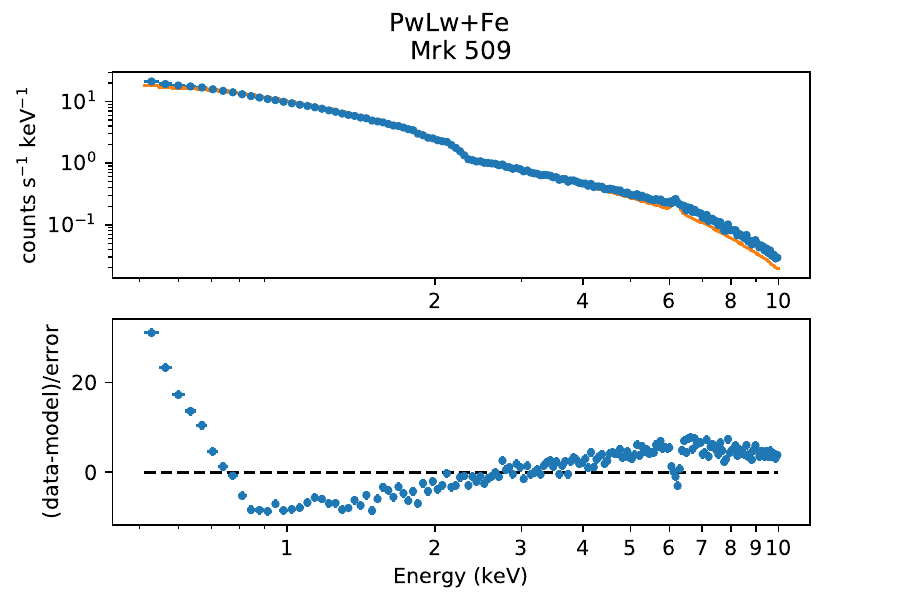} & \includegraphics[scale=0.65]{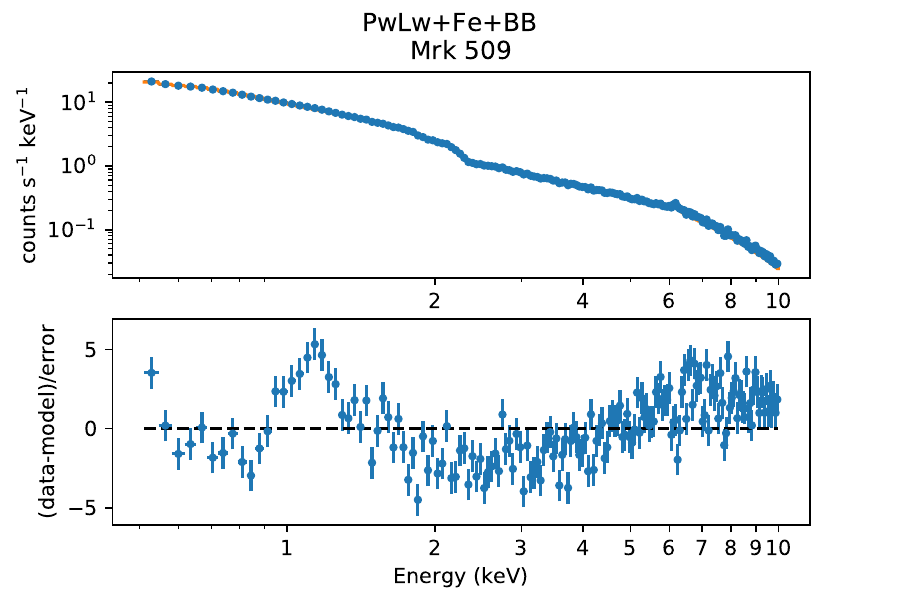} & \includegraphics[scale=0.65]{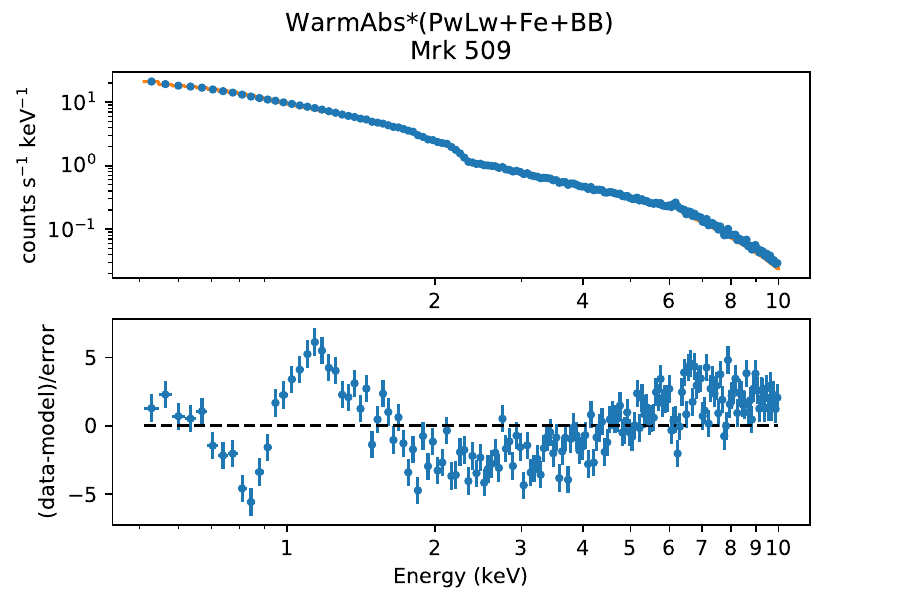} \\
    \includegraphics[scale=0.65]{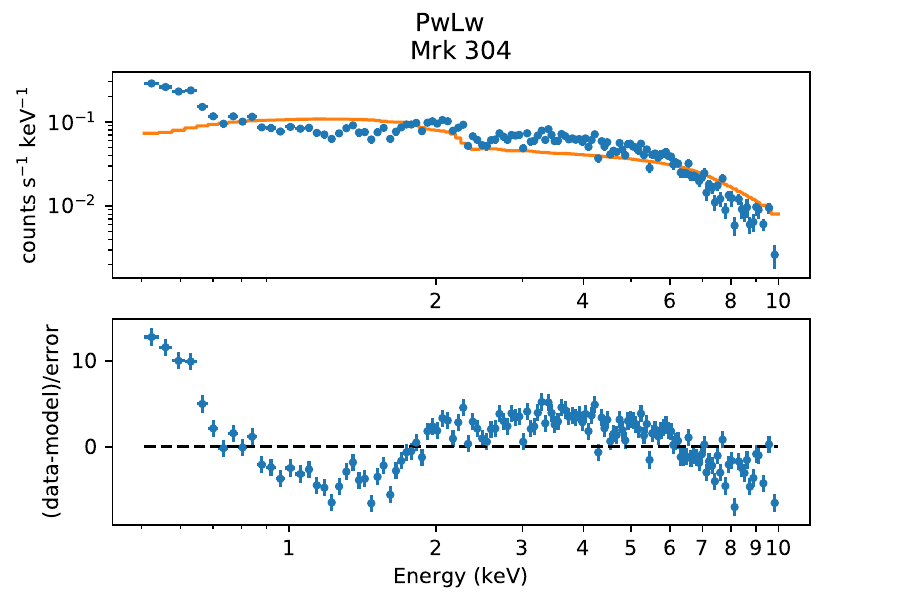} & \includegraphics[scale=0.65]{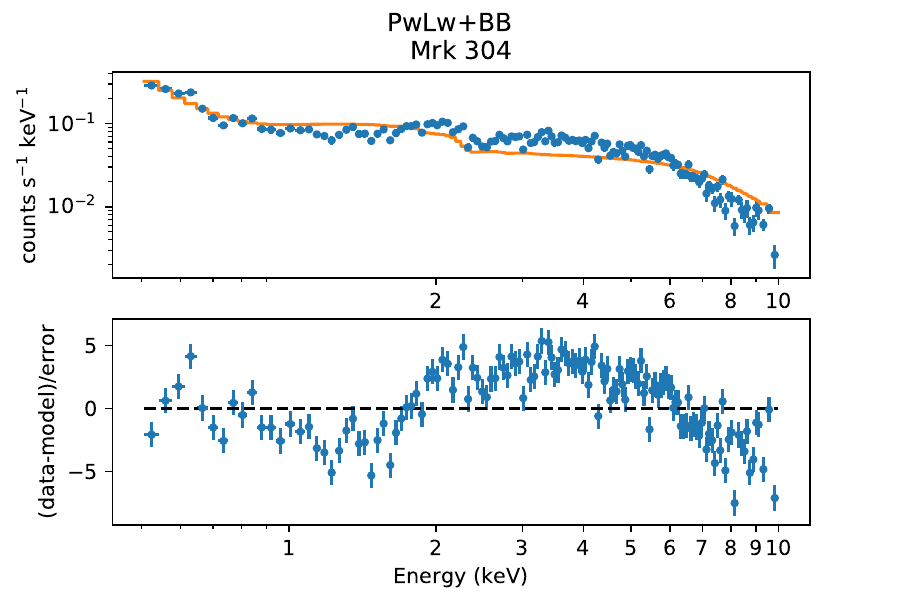} & \includegraphics[scale=0.65]{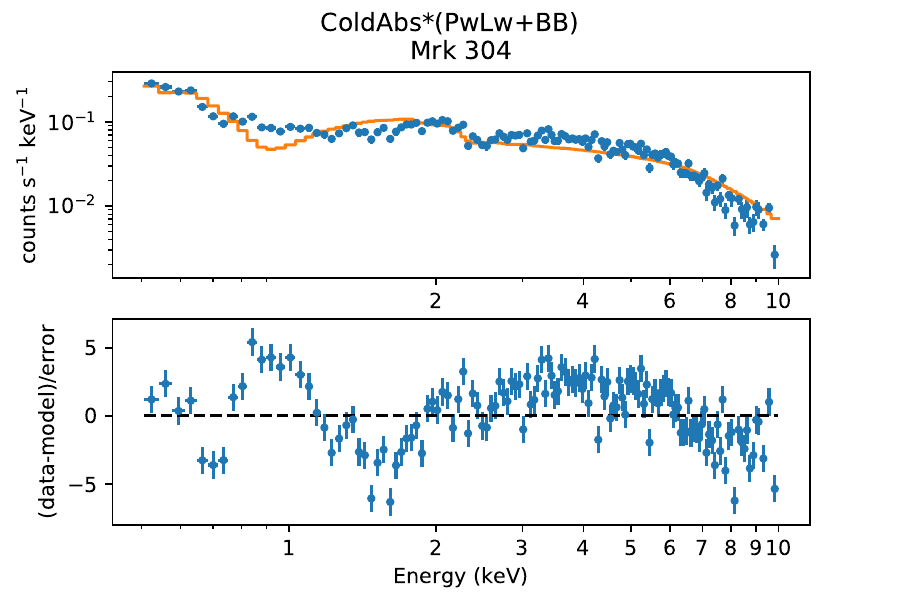} \\
\end{tabular}
\end{adjustbox}    
\caption{\textbf{Equatorial Polarized sources}. All the spectra are fitted in a range of 0.5-10 keV. First column corresponds to the baseline model: power law + significant Fe emission lines. Second column are fits with the addition of soft excess as black body. Third column is the resulting model of the absorption test, either cold, warm or not absorbed.}
\end{figure}
\end{document}